# Secure Routing in Wireless Mesh Networks

Jaydip Sen

*Innovation Lab, Tata Consultancy Services Ltd.*
*India*

## 1. Introduction

*Wireless mesh networks* (WMNs) have emerged as a promising concept to meet the challenges in next-generation networks such as providing flexible, adaptive, and reconfigurable architecture while offering cost-effective solutions to the service providers (Akyildiz et al., 2005). Unlike traditional Wi-Fi networks, with each *access point* (AP) connected to the wired network, in WMNs only a subset of the APs are required to be connected to the wired network. The APs that are connected to the wired network are called the *Internet gateways* (IGWs), while the APs that do not have wired connections are called the *mesh routers* (MRs). The MRs are connected to the IGWs using multi-hop communication. The IGWs provide access to conventional clients and interconnect ad hoc, sensor, cellular, and other networks to the Internet as shown in Fig. 1.

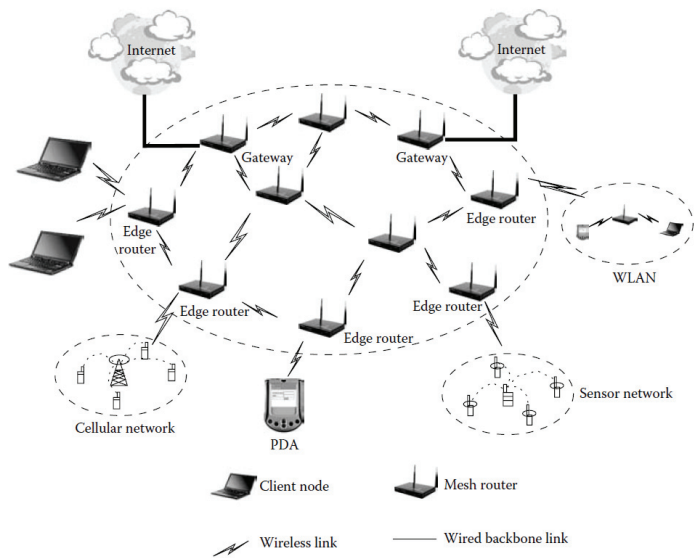

Fig. 1. The architecture of a wireless mesh network

Due to the recent research advances in WMNs, these networks have been used in numerous applications such as in home networking, community and neighborhood monitoring, security surveillance systems, disaster management and rescue operations etc (Franklin et al., 2007). As there is no wired infrastructure to deploy in the case of WMNs, they are considered cost-effective alternative to *wireless local area networks* (WLANs) and backbone networks to mobile clients. The existing wireless networking technologies such as IEEE 802.11, IEEE 802.15, IEEE 802.16, and IEEE 802.20 are used in the implementation of WMNs. As WMNs become an increasingly popular replacement technology for last-mile connectivity to the home networking, community and neighborhood networking, it is imperative to design an efficient resource management system for these networks. Routing is one of the most challenging issues in resource management for supporting real-time applications with stringent *quality of service* (QoS) requirements. However, most of the existing routing protocols for WMNs are extensions of protocols originally designed for *mobile ad hoc networks* (MANETs) and thus they perform sub-optimally. Moreover, most routing protocols for WMNs are designed without security issues in mind, where the nodes are all assumed to be honest. In practical deployment scenarios, this assumption does not hold. In a community-based WMN, a group of MRs managed by different operators form an access network to provide last-mile connectivity to the Internet. As with any end-user supported infrastructure, ubiquitous cooperative behavior in these networks cannot be assumed *a priori*. Preserving scarce access bandwidth and power, as well as security concerns may induce some selfish users to avoid forwarding data for other nodes, even as they send their own traffic through the network. The selfish behavior of an MR degrades the performance of a WMN since it increases the latency in packet delivery and packet drops and decreases the network throughput. In addition, some nodes may also launch malicious packet dropping attacks. Therefore, enforcing cooperation among the nodes in WMNs becomes a critical issue and a routing protocol should make use of such a cooperation enforcement scheme in order to ensure efficiency in packet forwarding and minimizing packet drops (Dong, 2009). To enforce cooperation among nodes and detect malicious and selfish nodes in self-organizing networks such as MANETs, various collaboration schemes have been proposed in the literature (Santhanam et al., 2008). Most of these proposals are based on trust and reputation frameworks which attempt to identify misbehaving nodes by an appropriate detection and decision making system, and then isolate or punish them. Unfortunately, most of these schemes are not directly applicable for WMNs due to inherent differences in characteristics between MANETs and WMNs. Efficient, reliable and secure routing protocols for WMNs are clearly in demand.

Keeping this in mind, this chapter provides a comprehensive overview of security issues in WMNs and then particularly focuses on secure routing in these networks. First, it identifies security vulnerabilities in the *medium access control* (MAC) and the network layers. Various possibilities of compromising data confidentiality, data integrity, replay attacks and offline cryptanalysis are also discussed. Then various types of attacks in the MAC and the network layers are discussed. In the MAC layer, attacks such as passive eavesdropping, link layer jamming (Law et al., 2005; Brown et al., 2006), MAC spoofing, replay attacks (Mishra et al., 2002) are discussed in detail. In the network layer, two broad categories of attacks are identified: (i) attacks on the control plane and (ii) attacks on the data plane. Among the attacks on the control plane, rushing attack (Hu et al., 2003a), wormhole attack (Hu et al., 2003b), blackhole attack (Al-Shurman et al., 2004), grayhole attack (Sen et al., 2007), Sybil attack (Newsome et al., 2004) are discussed. The data plane attacks are launched by the selfish and malicious nodes which lead to degradation in the network performance (Zhong et al., 2005; Salem et al., 2003). After enumerating the various types of attacks on the MAC and the network layer, the chapter briefly discusses on some of the preventive mechanisms for those attacks. After the preliminary discussion on various attacks and their countermeasures, the chapter focuses on its major issue- security in routing. It first identifies the major security requirements for design of a routing protocol in WMNs. Then various existing secure routing protocols for self-organizing networks such as ARAN (Sanzgiri et al., 2002), SAODV (Zapata et al., 2002), SRP (Papadimitratos et al., 2002), SEAD (Hu et al., 2002b), ARIADNE (Hu et al., 2002a), SEAODV (Li et al., 2011) etc. are discussed. All these protocols are compared in terms of their relative performance and their areas of application. After discussing these existing mechanisms, the chapter presents two novel secure routing protocols that detect selfish nodes in WMNs and isolate those nodes from the network activities so as to maximize the network throughput while providing desired QoS of the user application (Sen, 2010a; Sen, 2010b).

The organization of the chapter is as follows. In Section 2, we discuss various security vulnerabilities in different layers of the protocol stack of a WMN. Attacks at the physical, MAC, network, and transport layers are discussed in detail, and the countermeasures to defend against such attacks are briefly presented. In Section 3, several routing challenges in WMNs are highlighted. Section 4 presents some of the well-known existing security mechanisms for routing in WMNs. These protocols are also compared with respect to their capabilities in defending against different attacks in the network layer of WMNs. In Section 5, two novel routing protocols for WMNs are presented. These protocols can guarantee application QoS in addition to identifying malicious and selfish nodes in the network. Section 6 concludes the chapter while identifying some open issues and future research directions in designing secure routing protocols for WMNs.

In summary, the chapter makes the following contributions:

- It proposes threat models and security goals for secure routing in WMNs.
- It identifies various possible attacks on different layers of a WMN.
- It demonstrates how attacks against MANETs and peer-to-peer networks can be adapted into powerful attacks against WMNs.
- It makes security analysis of some of the major existing routing protocols fro WMNs.
- It presents various defense mechanisms to counter the well-known attacks on the routing protocols of WMNs.
- It presents two novel routing protocols for WMNs. These protocols enhance the routing efficiency and the application QoS while providing security in routing.
- It identifies some open research problems in the area of secure routing in WMNs.

## 2. Security Vulnerabilities in WMNs

Several vulnerabilities exist in the protocols foe WMNs. These vulnerabilities can be exploited by the attackers to degrade the performance of the network. The nodes in a WMN depend on the cooperation of the other nodes in the network. Consequently, the MAC layer and the network layer protocols for these networks usually assume that the participating nodes are honest and well-behaving with no malicious or dishonest intentions. In practice, however, some nodes in a WMN may behave in a selfish manner or may be compromised by malicious users. The assumed trust and the lack of accountability due to the absence of a central administrator make the MAC and the network layer protocols vulnerable to various types of attacks. In this section, a comprehensive discussion on various types of attacks in different layers of the protocol stack of a WMN is provided.

## 2.1 Physical layer attacks

The physical layer is responsible for frequency selection, carrier frequency generation, signal detection, modulation, and data encryption. As with any radio-based medium, the possibility of jamming attacks in this layer of WMNs is always there. Jamming is a type of attack which interferes with the radio frequencies that the nodes use in a WMN for communication (Shi et al., 2004). A jamming source may be powerful enough to disrupt communication in the entire network. Even with less powerful jamming sources, an adversary can potentially disrupt communication in the entire network by strategically distributing the jamming sources. An intermittent jamming source may also prove detrimental as some communications in WMNs may be time-sensitive. More complex forms of radio jamming attacks have been studied in (Xu et al., 2005), where the attacking devices do not obey the MAC layer protocols.

## 2.2 MAC layer attacks

Different types of attacks are possible in the MAC layer of a WMN. Some of the major attacks at this layer are: passive eavesdropping, jamming, MAC address spoofing, replay, unfairness in allocation, pre-computation and partial matching etc. These attacks are briefly described in this subsection.

i.   **Passive eavesdropping:** the broadcast nature of transmission of the wireless networks makes these networks prone to passive eavesdropping by the external attackers within the transmission range of the communicating nodes. Multi-hop wireless networks like WMNs are also prone to internal eavesdropping by the intermediate hops, whereby a malicious intermediate node may keep the copy of all the data that it forwards without the knowledge of any other nodes in the network. Although passive eavesdropping does not affect the network functionality directly, it leads to the compromise in data confidentiality and data integrity. Data encryption is generally employed using strong encryption keys to protect the confidentiality and integrity of data.

ii.  **Link layer jamming attack:** link layer attacks are more complex compared to blind physical layer jamming attacks. Rather than transmitting random bits constantly, the attacker may transmit regular MAC frame headers (no payload) on the transmission channel which conforms to the MAC protocol being used in the victim network (Law et al., 2005). Consequently, the legitimate nodes always find the channel busy and back off for a random period of time before sensing the channel again. This leads to the denial-of-service for the legitimate nodes and also enables the jamming node to conserve its energy. In addition to the MAC layer, jamming can also be used to exploit the network and transport layer protocols (Brown et al., 2006). Intelligent jamming is not a purely transmit activity. Sophisticated sensors are deployed, which detect and identify victim network activity, with a particular focus on the semantics of higher-layer protocols (e.g., AODV and TCP). Based on the observations of the sensors, the attackers can exploit the predictable timing behavior exhibited by higher-layer protocols and use offline analysis of packet sequences to maximize the potential gain for the jammer. These attacks can be effective even if encryption techniques such as *wired equivalent privacy* (WEP) and *WiFi protocol access* (WPA) have been employed. This is because the sensor that assists the jammer can still monitor the packet size, timing, and sequence to guide the jammer. Because these attacks are based on carefully exploiting protocol patterns and consistencies across size, timing and sequence, preventing them will require modifications to the protocol semantics so that these consistencies are removed wherever possible.

iii. **Intentional collision of frames:** a collision occurs when two nodes attempt to transmit on the same frequency simultaneously (Wood et al., 2002). When frames collide, they are discarded and need to be retransmitted. An adversary may strategically cause collisions in specific packets such as acknowledgment (ACK) control messages. A possible result of such collision is the costly exponential back-off. The adversary may simply violate the communication protocol and continuously transmit messages in an attempt to generate collisions. Repeated collisions can also be used by an attacker to cause resource exhaustion. For example a naïve MAC layer implementation may continuously attempt to retransmit the corrupted packets. Unless these retransmissions are detected early, the energy levels of the nodes would be exhausted quickly. An attacker may cause unfairness by intermittently using the MAC layer attacks. In this case, the adversary causes degradation of real-time applications running on other nodes by intermittently disrupting their frame transmissions.

iv. **MAC spoofing attack:** MAC addresses have long been used as the singularly unique layer-2 network identifiers in both wired and wireless LANs. MAC addresses which are globally unique have often been used as an authentication factor or as a unique identifier for granting varying levels of network privileges to a user. This is particularly common in 802.11 WiFi networks. However, today's MAC protocols (802.11) and network interface cards do not provide any safeguards that would prevent a potential attacker from modifying the source MAC address in its transmitted frames. On the contrary, there is often full support in the form of drivers from manufacturers, which makes this particularly easy. Modifying MAC addresses in transmitted frames is referred to as MAC spoofing, and can be used by attackers in a variety of ways. MAC spoofing enables the attacker to evade *intrusion detection systems* (IDSs) that are in place. Further, today's network administrators often use MAC addresses in access control lists. For example, only registered MAC addresses are allowed to connect to the access points. An attacker can easily eavesdrop on the network to determine the MAC addresses of legitimate devices. This enables the attacker to masquerade as a legitimate user and gain access to the network. An attacker can even inject a large number of bogus frames into the network to deplete the resources (in particular, bandwidth and energy), which may lead to denial of services for the legitimate nodes.

v. **Replay attack:** the replay attack, often known as the *man-in-the-middle* attack (Mishra et al., 2002), can be launched by external as well as internal nodes. An external malicious node (not a member of WMN) can eavesdrop on the broadcast communication between two nodes (*A* and *B*) in the network as shown in Fig. 2. It can then transmit legitimate messages at a later stage of time to gain access to the network resources. Generally, the authentication information is replayed where the attacker deceives a node (node *B* in Fig. 2) to believe that the attacker is a legitimate node (node *A* in Fig. 2). On a similar note, an internal malicious node, which is an intermediate hop between two communicating node, can keep a copy of all relayed data. It can then retransmit this data at a later point in time to gain the unauthorized access to the network resources.

vi. **Pre-computation and partial matching attack:** unlike the above-mentioned attacks, where MAC protocol vulnerabilities are exploited, these attacks exploit the vulnerabilities in the security mechanisms that are employed to secure the MAC layer of the network. Pre-computation and partial matching attacks exploit the cryptographic primitives that are used at MAC layer to secure the communication. In a pre-computation attack or *time memory trade-off attack* (TMTO), the attacker computes a large amount of information (key, plaintext, and respective ciphertext) and stores that information before launching the attack. When the actual transmission starts, the attacker uses the pre-computed information to speed up the cryptanalysis process. TMTO attacks are highly effective against a large number of cryptographic solutions. On the other hand, in a partial matching attack, the attacker has access to some (cipher text, plaintext) pairs, which in turn decreases the encryption key strength, and improves the chances of success of the brute force mechanisms. Partial matching attacks exploit the weak implementations of encryption algorithms. For example, the IEEE80.11i standard for MAC layer security in wireless networks is prone to the sensor hijacking attack and the man-in-the-middle attack that exploit the vulnerabilities in IEEE802.1X. DoS attacks on the four-way handshake procedure in IEEE 80.211i.

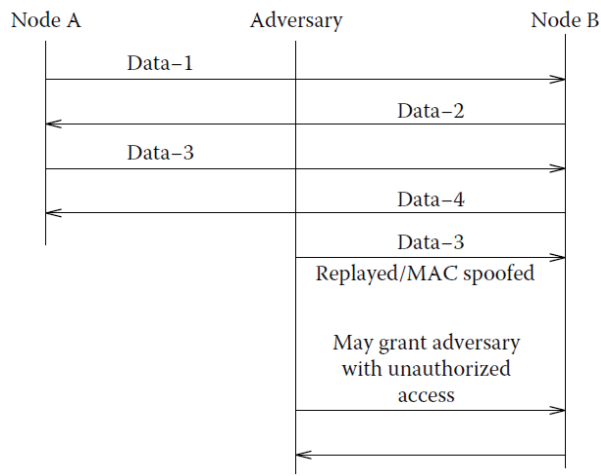

Fig. 2. Illustration of MAC spoofing and replay attacks

DoS attacks may also be launched by exploiting the security mechanisms. For example, the IEEE 802.11i standard for MAC layer security in wireless networks is prone to the sensor hijacking attack and the man-in-the-middle attack, exploiting the vulnerabilities in IEEE 802.1X, and DoS attack, exploiting vulnerabilities in the four-way handshake procedure in IEEEE 802.11i.

### 2.3 Network layer attacks

The attacks on the network layer can be divided into *control plane attacks* and *data plane attacks,* and can be active or passive in nature. Control plane attacks generally target the routing functionality of the network layer. The objective of the attacker is to make routes unavailable or force the network to choose sub-optimal routes. On the other hand, the data plane attacks affect the packet forwarding functionality of the network. The objective of the attacker is to cause the denial of service for the legitimate user by making user data undeliverable or injecting malicious data into the network. We first consider the network layer control plane attacks, and then the network layer data plane attacks.

**i.   Control plane attacks:** *Rushing* attacks (Hu et al., 2003a) targeting the on-demand routing
protocols (e.g., AODV) were among the first exposed attacks on the network layer of
multi-hop wireless networks. Rushing attacks exploit the route discovery mechanism of
on-demand routing protocols. In these protocols, the node requiring the route to the
destination floods the *route_request* (RREQ) message, which is identified by a sequence
number. To limit the flooding, each node only forwards the first message that it receives
and drops remaining messages with the same sequence number. To avoid collisions of the
messages, the protocol specifies a specific amount of delay between the receiving of a
route request message by a particular node, and its forwarding by the same node. The
malicious node launching the rushing attack forwards the RREQ message to the target
node before any other intermediate node from the source to destination. This can easily be
achieved by ignoring the specified delay. Consequently, the route from the source to the
destination includes the malicious node as an intermediate hop, which can then drop the
packets of the flow thereby launching a data plane DoS attack.

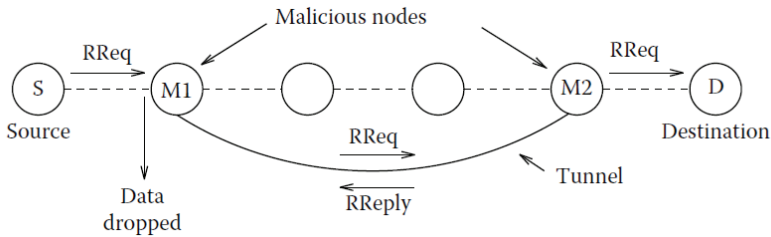

Fig. 3. Illustration of wormhole attack launched by nodes M1 and M2

A *wormhole* attack has a similar objective albeit it uses a different technique (Hu et al.,
2003b). During a wormhole attack, two or more malicious nodes collude together by
establishing a tunnel using an efficient communication medium (i.e., wired connection
or high-speed wireless connection etc.), as shown in Fig. 3. During the route discovery
phase of the on-demand routing protocols, the RREQ messages are forwarded between
the malicious nodes using the established tunnel. Therefore, the first RREQ message
that reaches the destination node is the one forwarded by the malicious nodes.
Consequently, the malicious nodes are added in the path from the source to the
destination. Once the malicious nodes are included in the routing path, these nodes
either drop all the packets resulting in a complete DoS attack, or drop the packets
selectively to avoid detection.

A *blackhole* attack (or *sinkhole* attack) (Al-Shurman et al., 2004) is another attack that
leads to denial of service in WMNs. It also exploits the route discovery mechanism of
on-demand routing protocols. In a blackhole attack, the malicious node always replies
positively to a RREQ, although it may not have a valid route to the destination. Because
the malicious node does not check its routing entries, it will always be the first to reply
to the RREQ message. Therefore, almost all the traffic within the neighborhood of the
malicious node will be directed towards the malicious node, which may drop all the
packets, resulting in denial of service. Fig. 4 shows the effect of a blackhole attack in the
neighborhood of the malicious node where the traffic is directed towards the malicious
node. A more complex form of the attack is the cooperative blackhole attack where multiple nodes collude together, resulting in complete disruption of routing and packet forwarding functionality of the network. The cooperative blackhole attack and the prevention mechanism have been studied in (Ramaswamy et al., 2003).

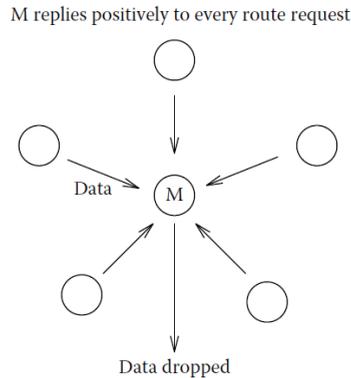

Fig. 4. Illustration of blackhole attack launched by node *M*

A *grayhole* attack is a variant of the blackhole attack (Sen et al., 2007). In a blackhole attack, the malicious node drops all the traffic that it is supposed to forward. This makes detection of the malicious node a relatively easier task. In a grayhole attack, the adversary avoids the detection by dropping the packets selectively. A grayhole does not lead to complete denial of service, but it may go undetected for a longer duration of time. This is because the malicious packet dropping may be considered congestion in the network, which also leads to selective packet loss.

A *Sybil* attack is the form of attack where a malicious node creates multiple identities in the network, each appearing as a legitimate node (Newsome et al., 2004). A Sybil attack was first exposed in distributed computing applications where the redundancy in the system was exploited by creating multiple identities and controlling considerable system resources. In the networking scenario, a number of services like packet forwarding, routing, and collaborative security mechanisms can be disrupted by the adversary using a Sybil attack. Following form of the attack affects the network layer of WMNs, which are supposed to take advantage of the path diversity in the network to increase the available bandwidth and reliability. If the malicious node creates multiple identities in the network, the legitimate nodes will assume these identities to be distinct nodes and will add these identities in the list of distinct paths available to a particular destination. When the packets are forwarded to these fake nodes, the malicious node that created the identities processes these packets. Consequently, all the distinct routing paths will pass through the malicious node. The malicious node may then launch any of the above-mentioned attacks. Even if no other attack is launched, the advantage of path diversity is diminished, resulting in degraded performance.

In addition to the above-mentioned attacks, the network layer of WMNs are also prone to various types of attack such as: *route request (RREQ) flooding attack*, *route reply (RREP) loop attack*, *route re-direction attack*, *fabrication attack*, *network partitioning* attack etc. RREQ flooding is one of the simplest attacks in which a malicious node tries to flood the entire network with RREQ message. As a consequence, this causes a large number of unnecessary broadcast communications resulting in energy drains and bandwidth wastage in the network. A *routing loop* is a path that goes through the same nodes over and over again. As a result, this kind of attack will deplete the resources of every node in the loop and will lead to isolation of the destination node.

Fig. 5 describes two instances where *route re-direction attack* has been launched by a malicious node *M*. In case *A*, the malicious node *M* tries to initiate the attack by modifying the mutable fields in the routing messages. These mutable fields include hop count, sequence numbers and other metric-related fields. The malicious node *M* could divert the traffic through itself by advertising a route to the destination with a larger *destination sequence number* (DSN) than the one it received from the destination. In case *B*, route re-direction attack may be launched by modifying the metric field in the AODV routing message, which is the hop-count field in this case. The malicious node *M* simply modifies the hop count field to zero in order to claim that it has a shorter path to the destination.

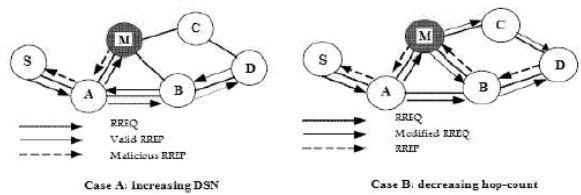

Fig. 5. Illustration of route re-direction attack

An adversary may fabricate false routing messages in order to disrupt routing in the network. For example, a malicious node may fabricate a *route error* (RERR) message in the AODV protocol. This may result in the upstream nodes re-initiating the route request to the unreachable destination so as to discover and establish alternative routes to them leading to energy and bandwidth wastage in the network. In a network partitioning attack, the malicious nodes collude together to disrupt the routing tables in such a way that the network is divided into disconnected partitions, resulting in denial of service for a certain network portion. Routing loop attacks affect the packet-forwarding capability of the network where the packets keep circulating in loop until they reach the maximum hop count, at which stage the packets are simply dropped.

ii. **Data plane attacks:** data plane attacks are primarily launched by selfish and malicious (compromised) nodes in the network and lead to performance degradation or denial of service of the legitimate user data traffic. The simplest of the data plane attacks is *passive eavesdropping*. Eavesdropping is a MAC layer attack. Selfish behavior of the participating WMN nodes is a major security issue because the WMN nodes are dependent on each other for data forwarding. The intermediate-hop selfish nodes may not perform the packet-forwarding functionality as per the protocol. The selfish node may drop all the data packets, resulting in complete denial of service, or it may drop the data packets selectively or randomly. It is hard to distinguish between such a selfish behavior and the link failure or network congestion. On the other hand, malicious intermediate-hop nodes may inject junk packets into the network. Considerable network resources (bandwidth and packet processing time) may be consumed to forward the junk packets, which may lead to denial of service for legitimate user traffic. The malicious nodes may also inject the maliciously crafted control packets, which may lead to the disruption of routing functionality. The control plane attacks are dependent on such maliciously crafted control packets. The malicious and selfish behaviors of nodes in WMNs have been studied in (Zhong et al., 2005; Salem et al., 2003).

## 2.4 Transport layer attacks

The attacks that can be launched on the transport layer of a WMN are flooding attack and de-synchronization attack. Whenever a protocol is required to maintain state at either end of a connection, it becomes vulnerable to memory exhaustion through flooding. An attacker may repeatedly make new connection request until the resources required by each connection are exhausted or reach a maximum limit. In either case, further legitimate requests will be ignored. De-synchronization refers to the disruption of an existing connection (Wood et al., 2002). An attacker may, for example, repeatedly spoof messages to an end host causing the host to request the retransmission of missed frames. If timed correctly, an attacker may degrade or even prevent the ability of the end hosts to successfully exchange data causing them instead to waste energy attempting to recover from errors which never really exist.

Table 1 presents various types of vulnerabilities in different layers of a WMN and their respective defense mechanisms.

| Layer | Attacks | Defense Mechanism |
|---|---|---|
| Physical | Jamming Device tampering | Spread-spectrum, priority messages, lower duty cycle, region mapping, mode change |
| MAC | Collision | Error-correction code |
| | Exhaustion | Rate limitation |
| | Unfairness | Small frames |
| Network | Spoofed routing information & selective forwarding | Egress filtering, authentication, monitoring |
| | Sinkhole | Redundancy checking |
| | Sybil | Authentication, monitoring, redundancy |
| | Wormhole | Authentication, probing |
| | Hello Flood | Authentication, packet leashes by using geographic and temporal info |
| | Ack. flooding | Authentication, bi-directional link authentication verification |
| Transport | Flooding De-synchronization | Client puzzles Authentication |
| Application | Logic errors Buffer overflow | Application authentication Trusted computing |

Table 1. Attacks on different layers of a WMN and their countermeasures

## 3. Routing Challenges in WMNs

In this section, some of the important challenges in designing routing protocols for WMNs are discussed. A typical architecture of a hierarchical WMN is presented in Fig. 1. At the top layer, are the *Internet gateways* (IGWs) which are connected to the wired Internet. They form the backbone infrastructure for providing Internet connectivity to the elements in the second level. The entities at the second level are called wireless *mesh routers* (MRs) that eliminate the need for wired infrastructure at every MR and forward their traffic in a multi-hop fashion towards the IGW. At the lowest level are the *mesh clients* (MCs) which are the wireless devices of the users. Internet connectivity and peer-to-peer communications inside the mesh are two important applications for a WMN. Therefore, design of an efficient and low-overhead routing protocol that avoids unreliable routes, and accurately estimate the end-to-end delay of a flow along the path from the source to the destination is a major challenge. Some of the major challenges in designing routing protocol for WMNs are discussed below:

i. **Measuring link reliability**: it has been observed that in wireless ad hoc networks like WMNs, nodes receiving broadcast messages introduce communication gray zones (Lundgren et al., 2002). In such zones, data messages cannot be exchanged although the hello messages reach the neighbors. This leads to disruption in communication among the nodes. Since the routing protocols such as AODV and WMR (Xue et al., 2003) relay on control packets like RREQ, these protocols are highly unreliable for estimating the quality of wireless links. Due to communication gray zone problem, nodes that are able to send and receive bi-directional RREQ packets sometimes cannot send/receive data packets at high rate. These fragile links trigger link repairs resulting in high control overhead.

ii. **End-to-end delay estimation:** an important issue in a routing protocol is end-to-end delay estimation. Current protocols estimate end-to-end delay by measuring the time taken to route route request (RREQ) and route reply (RREP) packets along the given path. However, RREQ and RREP packets are different from normal data packets and hence they are unlikely to experience the same levels of delay and loss as data packets. It has been observed through simulation that a RREP-based estimator overestimates while a hop-count-based estimator underestimates the actual delay experienced by the data packets (Kone et al., 2007). The reason for the significant deviation of a RREP-based estimator from the actual end-to-end delay is interference of signals. The RREQ packets are flooded in the network resulting in a heavy burst of traffic. This heavy traffic causes inter-flow interference in the paths. The unicast data packets do not cause such events. Moreover, as a stream of packets traverse along a route, due to the broadcast nature of wireless links, different packets in the same flow interfere with each other resulting in per-packet delays. Since the control packets do not experience per-packet delay, the estimates based on control packet delay deviate widely from the actual delay experience by the data packets.

iii. **Reduction of control overhead:** since the effective bandwidth of wireless channels vary continuously, reduction of control overhead is important in order to maximize throughput in the network. Reactive protocols such as AODV and DSR use flooding of RREQ packets for route discovery. This consumes a high proportion of the network bandwidth and reduces the effective throughput. An important challenge in designing a routing protocol for WMNs is to optimize the communication and computation overhead of the control messages so that the bandwidth of the wireless channels may be used for applications as efficiently as possible. Security and privacy issues bring another dimension of complexity. The goal of the protocol designer would be to design the security framework in such as way that it involves minimum computational and communication overhead.

## 4. Secure Routing Protocols for WMNs

Extensive work has been done in the area of secure unicast routing in multi-hop wireless networks (Hu et al., 2002a; Hu et al., 2002b; Sanzgiri et al., 2002; Marti et al., 2000; Papadimitratos et al., 2003a; Awerbuch et al., 2002; Awerbuch et al., 2005). As mentioned in Section 2.3, attacks on routing protocols can target either the route establishment process or the data delivery process, or both. Ariadne (Hu et al., 2002a) and SRP (Papadimitratos et al., 2003a) propose to secure on-demand source routing protocols by using hop-by-hop authentication techniques to prevent malicious packet manipulations on the route discovery process. SAODV (Zapata et al., 2002), SEAD (Hu et al., 2002b), and ARAN (Sanzgiri et al., 2002) propose to secure on-demand distance vector routing protocols by using one-way hash chains to secure the propagation of hop counts. The authors in (Papadimitratos et al., 2003b) propose a secure link state routing protocol that ensures the correctness of link state updates with digital signatures and one-way hash chains. To ensure correct data delivery, (Marti et al., 2000) proposes the *watchdog* and *pathrater* techniques to detect adversarial nodes by having each node monitor if its neighbors forward packets correctly. SMT (Papadimitratos et al., 2003a) and Ariadne (Hu et al., 2002a) use multi-hop routing to prevent malicious nodes from selectively dropping data. ODSBR (Awerbuch et al., 2002; Awerbuch et al., 2005) provides resilience to colluding Byzantine attacks by detecting malicious links based on end-to-end acknowledgment-based feedback technique. In HWMP (Bahr, 2006; Bahr, 2007), the on-demand node allows two mesh points (MPs) to communicate using peer-to-peer paths. This model is primarily used if nodes experience a changing environment and no root MP is configured. While the proactive tree building mode is an efficient choice for nodes in a fixed network topology, HWMP does not address security issues and is vulnerable to a numerous attacks such as RREQ flooding attack, RREP routing loop attack, route re-direction attack, fabrication attack, tunnelling attack etc (Li et al., 2011). LHAP (Zhu et al., 2003) is a lightweight transparent authentication protocol for wireless ad hoc networks. It uses TESLA (Perrig et al., 2000) to maintain the trust relationship among nodes, which is not realistic due to TESLA's delayed key disclosure period. In LHAP, simply attaching the TRAFFIC key right after the raw message is not secure since the traffic key has no relationship with the message being transmitted.

In contrast to secure unicast routing, work studying security problems specific to multicast routing in wireless networks is particularly scarce, with the notable exception of the work by (Roy et al., 2005) and BSMR (Curtmola et al., 2007). The work in (Roy et al., 2005) proposes an authentication framework that prevents outsider attacks in tree-based multicast protocol, MAODV (Royer et al., 2000), while BSMR (Curtmola et al., 2007) complements the work in (Roy et al., 2005) and presents a measurement-based technique that addresses insider attacks in tree-based multicast protocols.

A key point to note is that all of the above existing work in either secure unicast or multicast routing considers routing protocols that use only basic routing metrics, such as hop-count and latency. None of them consider routing protocols that incorporate high-throughput metrics, which have been shown to be critical for achieving high performance in wireless networks. On the contrary, many of them even have to remove important performance optimizations in existing protocols in order to prevent basic security attacks. There are also a few studies (Papadimitratos et al., 2006; Zhu et al., 2006) on secure QoS routing in wireless networks. However, they require strong assumptions, such as symmetric links, correct trust evaluation on nodes, ability to correctly determine link metrics despite attacks etc. In addition, none of them consider attacks on the data delivery phase. The work presented in (Dong, 2009) is the first of its kind that encompasses both high performance and security as goals in multicast routing and considers attacks on both path establishment and data delivery phases.

As mentioned in Section 2.3, wireless networks are also subject to attacks such as rushing attacks and wormhole attacks. Defenses against these attacks have been extensively studied in (Hu et al., 2003b; Hu et al., 2003a; Eriksson et al., 2006; Hu et al., 2004). RAP (Hu et al., 2003a) prevents the rushing attack by waiting for several flood requests and then randomly selecting one to forward, rather than always forwarding only the first one. Techniques to defend against wormhole attacks include *packet leashes* (Hu et al., 2003b) which restricts the maximum transmission distance by using time or location information. Truelink (Eriksson et al., 2006) which uses MAC level acknowledgments to infer whether a link exists between two nodes, and the work in (Hu et al., 2004) that relies on directional antennas are two mechanisms for defense against the wormhole attack.

In the following sub-sections, some of the well-known security protocols for routing in WMNs are presented. These protocols are extensions of base routing protocols like AODV, DSR etc. and use cryptographic mechanisms for ensuring node authentication, message integrity and message confidentiality.

## 4.1 Authenticated Routing for Ad Hoc Networks (ARAN)

*Authenticated routing for ad hoc networks* (ARAN) protocol (Sanzgiri et al., 2002), is an on-demand routing protocol that makes use of cryptographic certificates to offer routing security. It takes care of authentication, message integrity, and non-repudiation, but expects a small amount of prior security coordination among the nodes. In (Sanzgiri et al., 2002), vulnerabilities and attacks specific to AODV and DSR protocols are discussed and the two protocols are comapred with the ARAN protocol.

During the route discovery process of ARAN, the source node brodcasts *route_request* (RREQ) packets. The destination node, on receiving the RREQ packets, responds by unicasting back a reply packt, called the *route_reply* (RREP) packet. The ARAN protocol uses a preliminary cryptographic certification process, followed by an end-to-end route authentication process, which ensures secure route establishment. The protocol requires the use of a trusted certificate server $T$, whose public key is known to all the nodes in the network. End-to-end authentication is achieved by the source by having it verify that the intended destination was indeed reached. The source trusts the destination to choose the return path. The protocol is briefly discussed below.

**Issue of certificates:** ARAN utilizes an authenticated trusted server whose public key is known to all legitimate nodes in the network. The protocol assumes that keys are generated *a priori* by the server and distributed to all nodes in the network. It does not specify any specific key distribution algorithm. On joining the network, each node receives a certificate from the trusted server. The certificate received by a node $A$ from the trusted server $T$ looks like the following:

$$T \rightarrow A : cert_A = [IP_A, K_{A+}, t, e]K_{T-} \qquad (1)$$

In (1), $IP_A$, $K_{A+}$, $t$, $e$ and $K_{T-}$ represent the IP address of node $A$, the public key of node $A$, the time of creation of the certificate, the time of expiry of the certificate, and the private key of the server, respectively.

**End-to-end route authentication:** the main goal of the end-to-end route authentication process is to ensure that the packets reach the current intended destination from the source node. The source node *S* broadcasts a RREQ (i.e. route discovery) packet destined to the destination node *D*. The RREQ packet contains the packet identifier (*route discovery process* (RDP)), the IP address of the destination (*$IP_D$*), the certificate of the source node *S* (*$Cert_S$*), the current time (*t*) and a nonce $N_S$. The process can be denoted as in (2), where, $K_{S-}$ is the private key of the source node *S*.

$$S \rightarrow broadcasts := [RDP, IP_D, Cert_S, N_S, t]K_{S-} \tag{2}$$

Whenever the source sends a route discovery message, it increments the value of the nonce. Nonce is a counter used in conjunction with the time-stamp in order to make the nonce recycling easier. When a node receives an RDP packet from the source with a higher value of the source's nonce than that in the previously received RDP packets from the same source node, it makes a record of the neighbor from which it received the packet, encrypts the packet with its own certificate, and broadcasts it further. The process is represented in (3) below:

$$A \rightarrow broadcasts := [[RDP, IP_D, Cert_S, N_s, t]K_{s-}]K_{A-}, Cert_A \tag{3}$$

An intermediate node B on receiving an RDP packet from node A removes its neighbor's certificate, inserts its own certificate, and broadcast the packet further. The destination node, on receiving an RDP packet, verifies node *S*'s certificate and the tuple ($N_S$, *t*) and then replies with the *route reply* (REP). The destination unicasts the REP packet to the source node along the reverse path as in (4):

$$D \rightarrow X := [REP, IP_S, Cert_D, N_S, t]K_{D-} \tag{4}$$

In (4), node *X* is the neighbor of the destination node *D*, which had originally forwarded the RDP packet to node *D*. The REP packet follows the same procedure on the reverse path as that followed by the route-discovery packet. An error message is generated if the time-stamp or nonce does not match the requirements or if the certificate fails. The error message looks similar to the other packets except that the packet identifier is replaced by the ERR message.

In summary, ARAN is a robust protocol in the presence of attacks such as unauthorized participation, spoofed route signaling, fabricated routing messages, alteration of routing messages, securing shortest paths, and replay attacks. However, since ARAN uses public-key cryptography for authentication, it is particularly vulnerable to DoS attacks based on flooding the network with bogus control packets for which signature verifications are required. As long as a node can't verify signature at required speed, an attacker can force that node to discard some fraction of the control packets it receives.

## 4.2 Secure Efficient Ad Hoc Distance Vector (SEAD) routing protocol

*Secure efficient ad hoc distance vector* (SEAD) (Hu et al., 2002b) is a secure and proactive ad hoc routing protocol based on the *destination-sequenced distance vector* (DSDV) routing protocol (Perkins et al., 1994). This protocol is mainly designed to overcome security attacks such as DoS and resource consumption attacks. The operation of the routing protocol does not get affected even in the presence of multiple uncoordinated attackers corrupting the routing tables. The protocol uses a one-way hash function and does not involve any asymmetric cryptographic operation. The basic idea of SEAD is to authenticate the sequence number and metrics of a routing table update message using hash chain elements. The receiver also authenticates the sender ensuring that the routing information originates from the correct node. The source of each routing update message is also authenticated so as to prevent creation of a routing loop by an attacker launching an impersonation attack.

In the following, first a brief description of the base DSDV protocol is given followed by a discussion on the enhancements proposed in the SEAD protocol.

**Distance vector routing:** distance vector routing protocols belong to the category of table-driven routing protocols. Each node maintains a routing table containing the list of all known routes to various destination nodes in the network. The metric used for routing is the distance measured in terms of hop-count. The routing table is updated periodically by exchanging routing information. An alternative to this approach is *triggered updates*, in which each node broadcasts routing updates only if its routing table gets altered. The DSDV protocol for ad hoc wireless networks and WMNs uses *sequence number* tags to prevent the formation of loops, to counter the count-to-infinity problem, and for faster convergence. When a new route update packet is received for a destination, the node updates the corresponding entry in its routing table only if the sequence number on the received update is greater than that recorded with the corresponding entry in the routing table. If the received sequence number and the previously recorded sequence number are both equal, but if the routing update has a new value for the routing metric (distance in number of hops), then in this case also the update is effected. Otherwise, the received update packet is discarded. DSDV uses triggered updates (for important routing changes) in addition to the regular periodic updates. A slight variation of DSDV protocol known as *DSDV sequence number* (DSDV-SQ), initiates triggered updates on receiving a new sequence number update.

**One-way hash function**: SEAD uses authentication to differentiate between updates that are received from non-malicious nodes and malicious nodes. This minimizes the chances of resource consumption attacks caused by malicious nodes. SEAD uses a one-way hash function for authenticating the updates. A one-way hash function ($H$) generates a one-way hash chain ($h_1$, $h_2$, …). The function $H$ maps an input bit-string of any length to a fixed length bit-string, that is, $H : (0, 1)* \rightarrow (0, 1)^\rho$, where $\rho$ is the length in bits of the output bit-string. To create a one-way hash chain, a node generates a random number with initial value $x \in (0, 1)^\rho$. $h_0$, the first number in the hash chain is initialized to $x$. The remaining values in the chain are computed using the general formula $h_i = H(h_{i-1})$ for $0 \le i \le n$, for some $n$. The way one-way hash function incorporates security into the existing DSDV-DQ routing protocol will now be explained. The SEAD protocol assumes an upper bound on the metric used. For example, if the metric used is distance, then the upper bound value $m - 1$ defines the maximum diameter (maximum of lengths of all the routes between a pair of nodes) of the ad hoc wireless network or the WMN. Hence, the routing protocol assumes that no route of length greater than $m$ hops exists between any two nodes.

If the sequence of values calculated by a node using the hash function $H$ is given by ($h_1$, $h_2$… $h_n$), where $n$ is divisible by $m$, then for a routing table entry with sequence number $i$, let

$k = \dfrac{k}{m} - i$ . If the metric $j$ (distance) used for that routing table entry is, $0 \le j \le m - 1$ , then the

value of $h_{km+j}$ is used to authenticate the routing update entry for that sequence number $i$ and that metric $j$. Whenever a route update message is sent, the node appends the value used for authentication along with it. If the authentication value used is $h_{km+j}$, then the attacker who tries to modify this value can do so only if he/she knows $h_{km+j-1}$. Since it is a one-way hash chain, calculating $h_{km+j-1}$ becomes impossible. An intermediate node, on receiving this authenticated update, calculates the new hash value based on the earlier updates ($h_{km+j-1}$), the value of the metric, and the sequence number. If the calculated value matches with the one present in the route update message, then the update is done. Otherwise, the received update is just discarded.

SEAD avoids routing loops unless the loop contains more than one attacker. This protocol could be implemented easily with slight modifications to the DSDV protocol. The use of one-way hash chain to verify the authentication largely reduces the computational complexity. Moreover, the protocol is robust against multiple uncoordinated attacks. The main disadvantage is that a trusted entity is needed in the network to distribute and maintain the verification element of every node since the verification element of a hash chain is detached by a trusted entity. This leads to a single-point of failure in the protocol. If the trusted entity is compromised, the entire network becomes vulnerable. In addition, the protocol is vulnerable in situations where an attacker uses the same metric and sequence number which has been used in a recent update message and sends a new routing update.

### 4.3 Security-Aware Ad Hoc Routing (SAR) protocol

The *security-aware ad hoc routing* (SAR) protocol (Yi et al., 2001) uses security as one of the key metrics in path finding and provides a framework for enforcing and measuring the attributes of the security metric. This framework also enables the use of different levels of security for different applications that use SAR for routing. In WMNs, communication between two end nodes through possibly multiple nodes is based on the fact that the end nodes trust the intermediate nodes. SAR defines *level of trust* as a metric for routing and as one of the attributes for security to be taken into consideration. In SAR, security metric is embedded into the RREQ packet and the forwarding behavior of the protocol is implemented with respect to the RREQs. The intermediate nodes receive an RREQ packet with a particular security metric or trust level. The protocol ensures that a node can only process the packet or forward it if the node itself can provide the required security or has the required authorization or trust level. If the node cannot provide the required security, the RREQ is dropped. If an end-to-end path with the required security attributes can be found, a suitably modified RREP is sent from an intermediate node or the destination node. The routing protocol based on the level of trust is explained using Fig. 6.

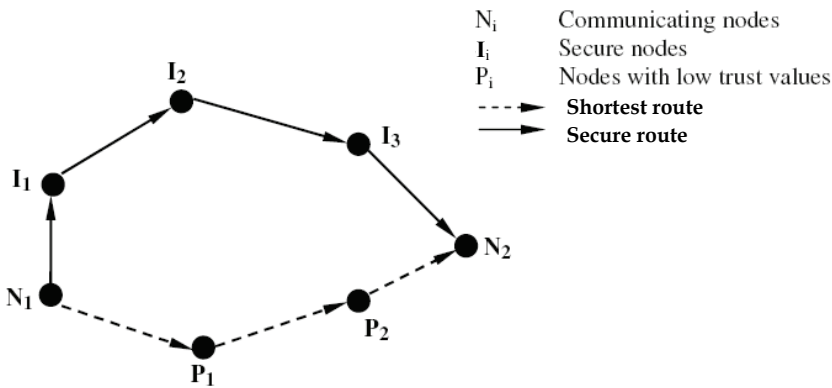

Fig. 6. Illustration of use of trust metric of nodes in routing

As shown in Fig. 6, two paths exist between the nodes $N_1$ and $N_2$ who want to communicate with each other. One of these paths is shorter which passes through private nodes ($P_1$ and $P_2$) whose trust levels are low. Hence, the protocol chooses a longer but secure path which passes through secure nodes $I_1$, $I_2$, and $I_3$.

The SAR protocol can be explained using any one of the traditional routing protocols. In this Section, SAR protocol has been explained using AODV protocol (Perkins et al., 1999). In the AODV protocol, the source node broadcasts a *route_request* (RREQ) packet to its neighbors. An intermediate node, on receiving a RREQ packet, forwards it further if it does not have a route to the destination. Otherwise, it initiates *route_reply* (RREP) packet back to the source node using the reverse path traversed by the RREQ packet. In SAR, a certain level of security is incorporated into the packet-forwarding mechanism. Here, each packet is associated with a security level which is determined by a number calculation method (explained later in this section). Each intermediate node is also associated with a certain level of security. On receiving a packet, the intermediate node is also associated with a certain level of security. On receiving a packet, the intermediate node compares its level of security with that defined for the packet. If node's security level is less than that of the packet, the RREQ is simply discarded. If it is greater, the node is considered to be a secure node and is permitted to forward the packet in addition to being able to view the packet. If the security level of the intermediate node and the received packet are found to be equal, then the intermediate node will not be able to view the packet (which can be ensured using a proper authentication mechanism); it just forwards the packet further.

Nodes of equal level of trust distribute a common key among themselves and with those nodes having higher levels of trust. Hence, a hierarchical level of security could be maintained. This ensures that an encrypted packet can be decrypted (using the common key) only by nodes of the same or higher levels of security compared to the level of security of the packet. Different levels of trust can be defined using a number calculated based on the level of security required. It can be calculated using a number of methods. Since timeliness, in-order delivery of packets, authenticity, authorization, integrity, confidentiality, and non-repudiation are some of the desired characteristics of a routing protocol, a suitable number can be defined for the trust level for nodes and packets based on the number of such characteristics taken into account.

The SAR protocol can be easily incorporated into the traditional routing protocols for ad hoc wireless networks and WMNs. It could be incorporated into both on-demand and table-driven routing protocols. The SAR protocol allows the application to choose the level of security it requires. But the protocol requires different keys for different levels of security. This tends to increase the number of keys required when the number of security levels used increases.

### 4.4 Secure Ad Hoc On-Demand Distance Vector (SAODV) routing protocol

In this section, a secure version of the AODV protocol will be described that plugs some well-known vulnerabilities of the routing protocol. Before presenting the secure version, a brief discussion of the base AODV protocol is presented.

**Ad hoc on-demand distance vector (AODV) routing protocol:** it is a reactive routing protocol (Perkins et al., 1999; Perkins et al., 2003) for MANETs and WMNs that maintains routes only between nodes which need to communicate. The routing messages do not contain information about the whole routing path, but only about the source and the destination. Therefore, routing messages do not have an increasing size. It uses destination sequence numbers to specify how fresh a route is (in comparison to the others), which is used to grant loop freedom.

Whenever a node needs to send a packet to a destination for which it has no 'fresh enough' route (i.e., a valid route entry for the destination whose associated sequence number is at least as great as the one contained in any RREQ that the node has received for that destination), it broadcasts an RREQ message to its neighbors. Each node that receives the broadcast message sets up a reverse route towards the originator of the RREQ, unless it has a 'fresher' one (Fig. 7). When the intended destination (or an intermediate node that has a 'fresh enough' route to the destination) receives the RREQ, it replies by sending an RREP. It is important that the only mutable information in an RREQ and in an RREP is the hop-count (which is being monotonically increased at each hop). The RREP is unicast back to the originator of the RREQ (Fig. 8).

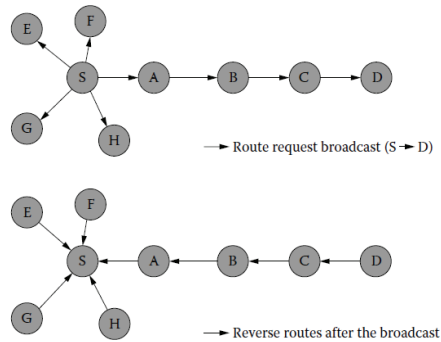

Fig. 7. Route request in AODV. *S* and *D* are the source and destination nodes respectively

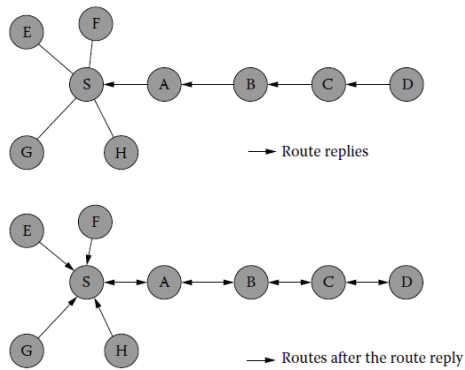

Fig. 8. Route reply in AODV. *S* and *D* are the source and destination nodes respectively

At each intermediate node, a route to the destination is set unless the node has a 'fresher' route than the one specified in the RREP). In the case that the RREQ is replied to by an intermediate node (and if the RREQ had set this option), the intermediate node also sends an RREP to the destination. In this way, it can be granted that the node path is being set up bi-directionally. In the case that a node receives a new route (by an RREQ or by an RREP) and the node already has a route 'as fresh' as the received one, the shortest one will be updated. Optionally, *route_reply acknowledgment* (RREP-ACK) message may be sent by the originator of the RREQ to acknowledge the receipt of the RREP. An RREP-ACK message has no mutable information. In addition to these routing messages, a *route_error* (RERR) message is used to notify the other nodes that certain nodes are not reachable anymore due to link breakage. When a node re-broadcasts an RERR, it only adds the unreachable destinations to which the node might forward messages. Therefore, the mutable information in an RERR is the list of unreachable destinations and the counter of unreachable destinations included in the message. It is predictable that, in each hop, the unreachable destination list may not change or become a subset of the original one.

Because AODV has no security mechanisms, malicious nodes can perform many attacks just by not following the protocol. A malicious node $M$ can carry out the following attacks (among many others) against AODV:

- Impersonate a node $S$ by forging an RREQ with its address as the originator address.
- When forwarding an RREQ generated by node $S$ to discover a route to node $D$, reduce the hop count field to increase the chances of being in the route path between $S$ and $D$ so that it can analyze the traffic between them.
- Impersonate a node $D$ by forging an RREP with its address as a destination address.
- Impersonate a node by forging an RREP that claims that the node is the destination.
- Selectively drop certain RREQs and RREPs and data packets. This kind of attack is especially hard even to detect because transmission errors have similar effect.
- Forge an RERR message pretending it is the node $S$ and send it to its neighbor $D$. The RERR message has a very high destination sequence number (*dsn*) for one of the unreachable destination, say, $U$. This might cause $D$ to update the destination sequence number corresponding to $U$ with the value *dsn* and, therefore, future route discoveries performed by $D$ to obtain a route to $U$ will fail (because $U$'s destination sequence number will be much smaller than the one stored in $D$'s routing table).
- According to the AODV specification (Perkins et al., 1999), the originator of an RREQ can put a much bigger destination sequence number than the real one. In addition, sequence numbers wrap around when they reach the maximum value allowed by the field size. This allows a very easy attack, where an attacker is able to set the sequence number of a node to any desired value by just sending two RREQ messages.

To plug these vulnerabilities the secure version of the AODV protocol is now presented.

**Secure ad hoc on-demand distance vector (SAODV) routing protocol:** this protocol has been proposed to secure the AODV protocol (Zapata et al. 2002). The idea behind SAODV is to use a signature to authenticate most of the fields of RREQs and RREPs and to use hash chains to authenticate the hop-count. SAODV designs signature extensions to AODV. Network nodes authenticate AODV routing packets with an SAODV signature extension, which prevents certain certain impersonation attacks. In SAODV, an RREQ packet includes a *route request single signature extension* (RREQ-SSE). The initiator chooses a maximum hop count, based on the expected network diameter, and generates a one-way hash chain of length equal to the maximum hop count plus one. This one-way hash chain is used as a metric authenticator, much like the hash chain within SEAD protocol (Hu et al., 2002b). The initiator signs the RREQ and the anchor of this hash chain; both this signature and the anchor are included in the RREQ-SSE. In addition, the RREQ-SSE includes an element of the hash chain based on the actual hop count in the RREQ header. For sake of explanation, we call this value the *hop-count authenticator* (HCA). For example, if the hash chain values $h_0$, $h_1$, ….., $h_N$ were generated such that $h_i = H[h_{i+1}]$, then the hop-count authenticator $h_i$ corresponds to a hop count of $N – i$.

With the exception of the hop-count field and HCA, the fields of the RREQ and RREQ-SSE headers are immutable and therefore can be authenticated by verifying the signature in the RREQ-SSE extension. To verify the hop-count field in the RREQ header, a node can follow the hash chain to the anchor. For example, if the hop-count field is $i$, then HCA should be $H^i[h_N]$. Because the length ($N$) and the anchor ($h_N$) of this hash chain are included in the RREQ-SSE and authenticated by the signature, a node can follow the hash chain and ensure that $h_N = H^{N-i}[HCA]$.

When forwarding an RREQ in SAODV, a node first authenticates the RREQ to ensure that each field is valid. It then performs duplicate suppression to ensure that it forwards only a single RREQ for each route discovery. The node then increments the hop-count field in the RREQ header, hashes the HCA, and re-broadcasts the RREQ, together with its RREQ-SSE extension. When the RREQ reaches the target, the target checks the authentication in the RREQ-SSE. If the RREQ is valid, the target returns an RREP as in AODV. A *route reply single signature extension* (RREP-SSE) provides authentication for the RREP. As in the RREQ, the only mutable field is the hop-count; as a result, the RREP is secured in the same way as the RREQ. In particular, an RRE-SSE has a signature covering the hash chain anchor together with all RREP fields except the hop count. The hop-count is authenticated by an HCA, which is also a hash chain element; an HCA $h_i$ corresponds to a hop-count of $N – i$.

A node forwarding an RREP checks the signature extension. If the signature is valid, then the forwarding node sets its routing table entry for the RREP's original source, specifying that packets to that destination should be forwarded to the node from which the forwarding node heard the RREP. For example, in Fig. 9, when node $B$ forwards the RREP from node $C$, it sets its next hop for destination node $D$ to $C$.

$$S \rightarrow * : \left\langle (RREQ, id, S, seq_S, D, oldseq_D, h_0, N)_{K_S^-}, o, h_N \right\rangle$$

$$A \rightarrow * : \left\langle (RREQ, id, S, seq_S, D, oldseq_D, h_0, N)_{K_S^-}, 1, h_{N-1} \right\rangle$$

$$B \rightarrow * : \left\langle (RREQ, id, S, seq_S, D, oldseq_D, h_0, N)_{K_S^-}, 2, h_{N-2} \right\rangle$$

$$C \rightarrow * : \left\langle (RREQ, id, S, seq_S, D, oldseq_D, h_0, N)_{K_S^-}, 3, h_{N-3} \right\rangle$$

$$D \rightarrow C : \left\langle (RREP, D, S, seq_D, S, lifetime, h_0', N)_{K_D^-}, o, h_N' \right\rangle$$

$$C \rightarrow B : \left\langle (RREP, D, S, seq_D, S, lifetime, h_0', N)_{K_D^-}, 1, h_{N-1}' \right\rangle$$

$$B \rightarrow A : \left\langle (RREP, D, S, seq_D, S, lifetime, h_0', N)_{K_D^-}, 2, h_{N-2}' \right\rangle$$

$$A \rightarrow S : \left\langle (RREP, D, S, seq_D, S, lifetime, h_0', N)_{K_D^-}, 3, h_{N-1}' \right\rangle$$

Fig. 9. Route discovery in SAODV protocol. Node $S$ is discovering a route to node $D$

SAODV allows replies from intermediate nodes through the use of a *route reply double signature extension* (RREP-DSE). An intermediate node replying to an RREQ includes an RREP-DSE. The idea here is that to establish a route to the destination, an intermediate node must have previously forwarded an RREP from the destination. If the intermediate node has stored the RREP and the signature, it can then return the same RREP if the sequence number in that RREP is greater than the sequence number specified in the RREQ. However, some of the fields of that RREP, in particular the life-time field, are no longer valid. As a result, a second signature, computed by the intermediate node, is used to authenticate this field.

To allow replies based on routing information from an RREQ packet, the initiator includes a signature suitable for an RREP packet through the use of an RREQ-DSE. Conceptually, the RREQ-DSE is an RREQ and RREP rolled into one packet. To reduce overhead, SAODV uses the observation that the RREQ and RREP fields substantially overlap. In particular, the RREQ-DSE needs to include some flags, a prefix size, and some reserved fields, together with a signature valid for an RREP using those values. When a node forwards an RREQ-DSE, it caches the route and the signature in the same way as if it had forwarded an RREP.

SAODV also uses signatures to protect the *route error* (RERR) message used in route maintenance. In SAODV, each node signs the RERR it transmits, whether it's originating the RERR or forwarding it. Nodes implementing SADOV don't change their destination sequence number information when receiving an RERR because the destination doesn't authenticate the destination sequence number. Fig. 10 shows an example of SAODV route maintenance.

$$B \rightarrow A : (RERR, D, seq_D)_{K_B^-}$$

$$A \rightarrow S : (RERR, D, seq_D)_{K_A^-}$$

Fig. 10.  Route maintenance in SAODV protocol.

## 4.5 Secure Routing Protocol (SRP)

Papadimitratos et al. (Papadimitratos et al., 2002) have proposed a *secure routing protocol* (SRP) that can be applied to several existing routing protocols (in particular to DSR (Johnson et al., 2007)). It is an on-demand source routing protocol that captures the basic features of reactive routing. The packets in SRP have extension headers that are attached to RREQ and RREP messages. The protocol doesn't attempt to secure RERR packets; instead it delegates the route-maintenance function of the secure route maintenance portion of the *secure message transmission protocol*. SRP uses a sequence number in the RREQs and RREPs to ensure freshness, but this sequence number can only be checked at the target. SRP requires a security association only between communicating nodes and uses this security association to authenticate RREQs and RREPs computed through the use of *message authentication codes* (MACs). At the target, SRP can detect any modifications of the RREQs, and at the source node, it can detect modifications of the RREPs. In the following, the protocol is discussed briefly.

In SRP, *route requests* (RREQs) generated by a source node $S$ are protected by *message authentication codes* (MACs) computed using a key shared with the target $T$. Requests are broadcast to all the neighbors of $S$. Each neighbor that receives a request for the first time appends its identifier to the request and re-broadcasts it. The intermediate nodes also perform the same actions.  The MAC in the request is not checked because only $S$ and $T$ know the key being used to compute it. When the request reaches the target $T$, its MAC is checked by $T$. If it is valid, then it is assumed by the target that all adjacent pairs of nodes on the path of the RREQ are neighbors. Such paths are called valid or plausible routes. The target $T$ replaces the MAC of a valid RREQ by a MAC computed with the same key that authenticates the route. This is then sent back (upstream) to $S$ using the reverse route. For example, an RREQ that reaches an intermediate node $X_j$ is of the following form:

$$msg_{S,T,rreq} = (rreq, S, T, id, sn, X_1, X_2.........X_j, mac_S) \tag{5}$$

In (5), $id$ is a randomly generated route identifier, $sn$ is a session number and $mac_S$ is a MAC on *(rreq, S, T, id, sn)* computed by $S$ using a key shared with $T$, $X_1$, …….., $X_p$, $T$ is a discovered route, then the *route reply* (RREP) of the target $T$ has the following form for all intermediate nodes $X_j$, $1 \le j \le p$:

$$msg_{S,T,rrep} = (rrep, S, T, id, sn, X_1, X_2, ......X_p, mac_T) \tag{6}$$

In (6), $mac_T$ is a MAC computed by $T$ with the key shared with $S$ on the message field preceding it. Intermediate nodes should check the RREP header (including its *id* and *sn*) and that they are adjacent with two of their neighbors on the route before sending the RREP upstream.

SRP doesn't attempt to prevent unauthorized modification of fields that are ordinarily modified in the course of forwarding these packets. For example, a node can freely remove or corrupt the node list of an RREQ packet that it forwards. Since SRP requires a security association between communicating nodes, it uses extremely lightweight mechanisms to prevent other attacks. For example, to limit flooding, nodes record the rate at which each neighbor forwards the RREQ packets and gives priority to REQUEST packets sent through neighbor that less frequently forward REQUEST packets. Such mechanisms can secure a protocol when few attackers are present. However, such techniques provide secondary attacks, such as sending forged RREQ packets to reduce the effectiveness of a node's authentic RREQs. In addition, such techniques exacerbate the problem of greedy nodes. For example, a node that doesn't forward RREQ packets ordinarily achieves better performance because it is generally less congested, and it doesn't need to use its battery power to forward packets originated by other nodes. In SRP, a greedy node retains these advantages, and in addition, gets a higher priority when it initiates route discovery.

## 4.6 ARIADNE: A secure on-demand routing protocol for ad hoc networks

Ariadne (Hu et al., 2002a) is a secure on-demand routing protocol based on the *dynamic source routing* (DSR) protocol (Johnson et al., 2007). The protocol can withstand node compromise and relies only on highly efficient symmetric key cryptography. Ariadne can authenticate routing message using one of the three schemes: (i) shared secret between each pair of nodes, (ii) shared secrets between communicating nodes combined with broadcast authentication using TESLA (Perrig et al., 2001), and (iii) digital signatures. In this section, we discuss Ariadne with TESLA, an efficient broadcast authentication scheme that requires loose time synchronization. Using pair-wise shared keys the protocol avoids the need for time synchronization but at the cost of higher key-setup overhead. Ariadne discovers routes in a reactive (on-demand) manner through route discovery and uses them to source route data packets to their destinations. Each forwarding node helps by performing route maintenance to discover problems with each selected route.

**Route discovery:** The protocol design is explained in two stages: (i) a mechanism is presented that lets the target node verify the authenticity of the RREQ, and (ii) an efficient per-hop hashing technique is described that verifies whether any node is missed from the node list in the RREQ. In the following, we assume that the initiator node $S$ performs a route discovery for target node $D$ and that they share the secret keys $K_{SD}$ and $K_{DS}$, respectively for message authentication in each direction.

**i.** *Target authenticates route request*: To convince the target of the legitimacy of each field in an RREQ, the initiator simply includes a *message authentication code* (MAC) computed with the key $K_{SD}$ over unique data – for example, a timestamp. The target can easily verify the route requestor's authenticity and freshness using the shared key $K_{SD}$. In a route discovery, the initiator wants to authenticate each individual node in the node list of the RREP. A secondary requirement is that the target can authenticate each node in the node list of the RREQ so that it will return an RREP only along paths that contain legitimate nodes. Each hop authenticates the new information in the RREQ using its current TESLA key. The target node buffers the RREP until intermediate nodes can release the corresponding TESLA keys. The TESLA security condition is verified at the target node, and the target includes a MAC in the RREP to certify that security condition was met.

**ii.** *Per-hop hashing*: Authenticating data in routing messages isn't sufficient because an attacker could remove a node from the node list in an RREQ. One-way hash functions are used to verify that no hop was omitted – an approach that is called *per-hop hashing*. To change or remove a previous hop, an attacker must either hear an RREQ without that node listed or must be able to invert the one-way hash function. For efficiency, the authenticator may be included in the hash value passed in the RREQ. Fig. 11 shows an example of Ariadne route discovery.

$$S : h_0 = MAC_{K_{SD}}(REQUEST, S, D, id, ti)$$

$$S \rightarrow * : \langle REQUEST, S, D, id, ti, h_0, (), () \rangle$$

$$A : h_1 = H[A, h_0], \langle M_A = MAC_{K_{Ati}}(REQUEST, S, D, id, ti, h_1, (A), ()) \rangle$$

$$A \rightarrow * : \langle REQUEST, S, D, id, ti, \boldsymbol{h_1}, (\boldsymbol{A}), \boldsymbol{M_A} \rangle$$

$$B : h_2 = H[B, h_1], \langle M_B = MAC_{K_{Bti}}(REQUEST, S, D, id, ti, h_2, (A, B), (M_A)) \rangle$$

$$B \rightarrow * : \langle REQUEST, S, D, id, ti, \boldsymbol{h_2}, (A, \boldsymbol{B}), (M_A, \boldsymbol{M_B}) \rangle$$

$$C : h_3 = H[C, h_2], \langle M_C = MAC_{K_{Cti}}(REQUEST, S, D, id, ti, h_3, (A, B, C), (M_A, M_B)) \rangle$$

$$C \rightarrow * : \langle REQUEST, S, D, id, ti, \boldsymbol{h_3}, (A, B, \boldsymbol{C}), (M_A, M_B, \boldsymbol{M_C}) \rangle$$

$$D : M_D = \langle MAC_{K_{DS}}(REPLY, D, S, ti, (A, B, C), (M_A, M_B, M_C)) \rangle$$

$$D \rightarrow C : \langle REPLY, D, S, ti, (A, B, C), (M_A, M_B, M_C), \boldsymbol{M_D}, () \rangle$$

$$C \rightarrow B : \langle REPLY, D, S, ti, (A, B, C), (M_A, M_B, M_C), M_D, (\boldsymbol{K_{Cti}}) \rangle$$

$$B \rightarrow A : \langle REPLY, D, S, ti, (A, B, C), (M_A, M_B, M_C), M_D, (K_{Cti}, \boldsymbol{K_{Bti}}) \rangle$$

$$A \rightarrow S : \langle REPLY, D, S, ti, (A, B, C), (M_A, M_B, M_C), M_D, (K_{Cti}, K_{Bti}, \boldsymbol{K_{Ati}}) \rangle$$

Fig. 11. Route discovery in Ariadne. Initiator $S$ attempts to discover a route to target $D$. The bold font indicates changed message fields relative to the previous similar message.

**Route maintenance**: Route maintenance in Ariadne is based on the DSR protocol. A node forwarding a packet to the next hop along the source route returns an RERR to the packet's original sender if it is unable to deliver the packet to the next-hop after a limited number of retransmission attempts. The mechanisms for securing RERRs are discussed in the following. However, the case in which attackers to not send the RERRs is not considered. To prevent unauthorized nodes from sending RERRs, a mechanism should be in place in which the sender needs to authenticate the RERR messages. Each node on the return path to the source node forwards the RERR message. If the authentication is delayed – for example, when TESLA is used – each node that will be able to authenticate the RERR message buffers it until it can be authenticated.

**Avoiding routing misbehavior**: Ariadne protocol described above is vulnerable to an attacker that happens to be along the discovered route. In particular, a mechanism should be there that is able to determine whether the intermediate nodes forward the packets that they are requested to forward. To avoid the continued use of malicious routes, the routes are chosen based on their prior performance in packet forwarding. The scheme relies on feedback about which packets were successfully delivered. The feedback can be received either through an extra end-to-end network layer message or by exploiting properties of the transport layers, such as TCP with *selective acknowledgments* (Mathis et al., 1996). This feedback approach is somewhat similar to the one used in IPv6 for neighbor unreachability detection (Narten et al., 2007). A node with multiple routes to a single destination can assign a fraction of packets that it originates to be sent along each route. When a substantially smaller fraction of packets sent along any particular route is successfully delivered, the node can begin sending a smaller fraction of its overall packets to that destination along that route.

## 4.7 Security Enhanced AODV protocol

A *security enhanced AODV* (SEAODV) routing protocol has been proposed in (Li et al., 2011) that employs Blom's key pre-distribution scheme (Blom, 1985) to compute the *pair-wise transient key* (PTK) through the flooding of enhanced *hello* message and subsequently uses the established PTK to distribute the *group transient key* (GTK). PTK and GTK are used for authenticating unicast and broadcast routing messages respectively. In WMNs, a unique PTK is shared by each pair of nodes, while GTK is shared secretly between the node and all its one-hop neighbors. A *message authentication code* (MAC) is attached as the extension to the original AODV routing message to guarantee the message's authenticity and integrity in a hop-by-hop fashion. Since SEAODV uses Blom's key pre-distribution scheme, for the benefit of the readers, a brief discussion on the key pre-distribution scheme is presented in the following before the secure routing protocol is discussed.

**Blom's key pre-distribution scheme:** Blom's key pre-distribution is applied for implementing key exchange process (Blom, 1985; Du et al., 2003). Blom's $t$ secure key pre-distribution scheme is as follows. Blom's pre-distribution scheme is based on $(N, t + 1)$ *maximum distance separable* (MDS) linear codes (MacWilliams et al., 1977). In this scheme, before a network is deployed, a central authority first constructs a $(t + 1)$ x $N$ public matrix $P$ over a finite field $GF(q)$, where $N$ is the network size. Then, the central authority selects a random $(t + 1)$ x $(t + 1)$ symmetric matrix $S$ over $GF(q)$, where $S$ is secret and only known to the central authority. An $N$ x $(t + 1)$ matrix $A = (S . P)^T$ is computed, where $(.)^T$ denotes the transpose operator. The central authority pre-loads the $i$-th row and $i$-th column of $P$ to node $i$, for $i = 1, 2,…..n$. When node $i$ and $j$ need to establish a shared key, they first exchange their columns of $P$, and then node $i$ computes a key $K_{ij}$ as the product of its own row of $A$ and $j$-th column of $P$, and node $j$ computes $K_{ji}$ as the product of its own row of $A$ and the $i$-th column of $P$. Since $S$ is symmetric, it is easy to see that:

$$K = A \cdot P = (S \cdot P)^T \cdot P = P^T \cdot S^T \cdot P = P^T \cdot S \cdot P = (A \cdot P)^T = K^T \qquad (7)$$

The node pair $(i, j)$ uses $K_{ij} = K_{ji}$ as the shared key. The Blom scheme has a *t-secure property*. It implies that in a network of $N$ nodes, the collusion of less than $t +1$ nodes cannot reveal any key shared by other pairs of nodes. This is because as least $t$ rows of $A$ and $t$ columns of $P$ are required to solve the secret symmetric matrix $S$. The memory cost per node in the Blom scheme is $t + 1$. To guarantee perfect security in a WMN with $N$ nodes, the $(N - 2)$-secure Blom scheme should be used, which means the memory cost per node is $N - 1$. Hence Blom scheme can provide strong security in networks of small size.

**SEAODV protocol:** SEAODV is built on AODV protocol. It requires each node in the network to maintain two key hierarchies. One is the broadcast key hierarchy, which includes all the broadcast keys from its active one hop neighbors. The other hierarchy is called unicast hierarchy, which stores all secret pair-wise keys that this node shares with its one hop neighbors. Every node uses keys in its broadcast routing messages (e.g., RREQ messages) from its one hop neighbors and applies secret pair-wise keys in the unicast hierarchy to verify the incoming messages, such as the RREP messages. Various features of the protocol are now described.

i.   **Enhanced hello messages:** in AODV, hello message is broadcast by each node in its one-hop neighborhood. In SEAODV, two *enhanced hello* messages are defined following the idea presented in (Jing et al., 2004). Each node embeds its column of the public matrix $P$ into its enhanced *hello RREQ* message. Since each column of $P$ can be regenerated by applying the seed (a primitive element of $GF(q)$) from each node, every node only needs to store the seed in order to exchange the public information of matrix $P$. To guarantee bi-directional links, the neighboring nodes who receive *hello RREQ* reply with an enhanced *hello RREP*.

ii.  **Exchange public *Seed_P* and GTK using enhanced hello message:** during the key pre-distribution phase, every legitimate node in the WMN knows and stores the public *Seed_P* (seed of the column of public matrix $P$) and the corresponding private row of the generated matrix $A$. The entire exchange process is depicted in three steps: (a) exchange of Seed_P of public matrix $P$, (b) derivation of PTK, and (c) exchange of GTK. In the *exchange of Seed_P phase*, each node looks for its public *Seed_P* from its key pool, and broadcasts the enhanced *hello RREQ* message. On completion of this step, each node in the network possesses the public *Seed_P* of all of its one-hop neighbors. In the *derivation of PTK phase*, each node uses the *Seed_P* it received from its neighbors and the node's corresponding private row of matrix $A$ to compute PTK. On completion of this step, every node has stored the public *Seed_P* of its neighbors and has derived the PTK it shares with each of its one-hop neighbors. In the *exchange of GTK phase*, upon receiving hello RREQ from node $X$, node $Y$ (node $X$'s neighbor) encrypts *GTK_Y* with its private *PTK_Y* and unicasts the corresponding *hello RREP* message back to $X$. The encrypted *GTK_Y* is also attached in the unicast *hello RREP* message. Once $X$ receives *hello RREP* from $Y$, $X$ applies its private *PTK_X* to decrypt the *GTK_Y* and stores it in the database. The same process applies to node $Y$ as well. Eventually, every node possesses the GTK keys from all its one-hop neighbors and the group of secret pair-wise PTK keys that it shares with each of its one-hop neighbor.

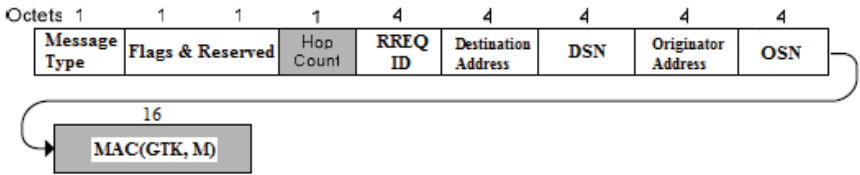

Fig. 12. The structure of RREQ message in SEAODV protocol

**iii. Securing route discovery:** in order to ensure hop-by-hop authentication, each node must verify the incoming message from its one-hop neighbors before re-broadcasting or unicasting the messages. The trust relationship between each pair of nodes relies on the shared GTK and PTK of the nodes. Route discovery process of SEAODV is similar to that of AODV, except for a MAC extension appended to the AODV message. The structure of the RREQ in SEAODV is presented in Fig. 12. The MAC is computed for message $M$ using GTK of the node which needs to broadcast a RREQ to its one-hop neighbors. When a node wants to discover a route to a designated destination, it broadcasts the modified RREQ message to its neighbors. The receiving node computes the corresponding MAC value of the received message if the node possesses the GTK of the sender. The receiving node then compares the computed MAC with the one it received. If there is a match, the received RREQ is considered to be authentic and unaltered. The receiving node then updates the mutable field (hop-count in RREQ) and its routing table, and subsequently sets up the reverse path back to the source by recording the neighbor from which it received the RREQ. Finally, the node computes a MAC of the updated RREQ with its GTK and attaches the MAC value to the end of the RREQ before the message is re-broadcast to its neighbors.

**iv. Securing route setup:** the destination node or an intermediate node generates a modified RREP and unicasts it back to the next hop from which it received the RREQ. Since the RREP message is authenticated at each hop using PTKs, an adversary has no opportunity to re-direct the traffic. Before unicasting the modified RREP back to the originator of the RREQ, the node first needs to check its routing table to identify the next hop from which it received the broadcast RREQ. The node then applies PTK that it shares with the identified next hop to compute the $MAC\ (PTK, M)$ and affixes this MAC to the end of RREP as shown in Fig. 13.

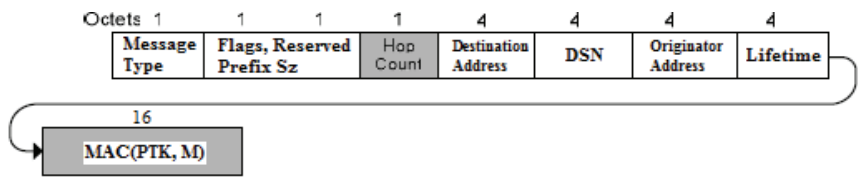

Fig. 13. The structure of RREP message in SEAODV protocol

Upon receiving the RREP from node $Y$, node $X$ checks whether $PTK\_YX$ is in its group PTK. If it is, then node $X$ computes $MAC'(PTK\_XY, M)$ and compares it with the $MAC(PTK\_YX, M)$ it received from node $Y$. If $MAC'(PTK\_XY, M)$ matches $MAC(PTK\_YX, M)$, the received RREP is considered authentic. Node $X$ then updates the hop-count field in the RREP and its own routing table, sets up the forwarding path towards the destination. Node $X$ also searches the appropriate PTK that it shares with its next hop to which the new RREP is going to be forwarded to the source. Node *X* then uses the PTK to construct the new MAC and appends it to the new RREP message. Otherwise, the received RREP is deemed to be unauthentic and hence dropped.

v.  **Securing route maintenance:** a node generates an RERR message if it receives data packet destined to another node for which it does not have an active route in its routing table or the node detects a broken link for the next hop of an active route or a node receives a RERR message from a neighbor for one or more active routes. The structure of a modified RERR message is presented in Fig. 14. The MAC field in the modified RERR message is computed by applying the node's GTK on the entire RERR packet. On receiving the broadcast RERR message from node *Y*, node *X* first checks whether it has the *GTK_Y*. If it has, node *X* then computes *MAC'(GTK_Y, M')* and compares it with the received MAC. If the two MACs match, node *X* searches its routing table and tries to identify the affected routes (a new group of unreachable destinations) that use node *Y* as its next-hop based on the unreachable destination list received from *Y*. If no routes in node *X*'s routing table is affected, *X* simply drops the RERR message and starts listening to the channel again. Node *X* also discards the RERR message if it fails to find the *GTK_Y* or the *MAC'(GTK_Y, M')* does not match the one received from node *Y*.

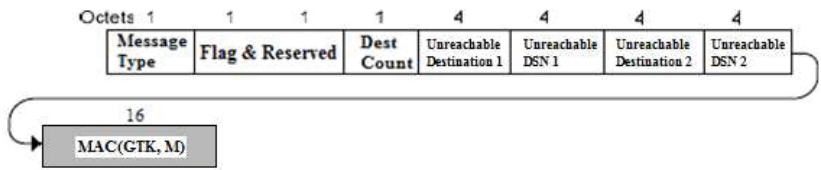

Fig. 14. The structure of RERR message in SEAODV protocol

**Security analysis of SEAODV:** SEAODV is vulnerable to RREQ flooding attack. However, since it authenticates RREQs from nodes that are in the list of active one-hop neighbors, the detection of the attack will be fast. Since GTKs and PTKs are used to secure the broadcast and unicast messages, and integrity of the messages are protected by MACs, the protocol is robust against RREP routing loop attack and route re-direction attack. RERR fabrication attack has minimal impact on SEAODV protocol, since a receiving node authenticates RERR messages coming from its active one-hop neighbors only. Since a malicious node can only forward the replayed RERR messages coming from the receiving node's one-hop neighbors, launching of RERR fabrication attack becomes particularly difficult.

## 5. Some novel secure routing protocols for WMNs

In this section, two novel routing protocols for WMNs are presented that can satisfy application QoS requirements in addition to providing security in routing. The first protocol is based on a reliable estimation of the available bandwidth in wireless links and a robust estimation of the end-to-end delay on a routing path. The protocol, while satisfying the application QoS, detects selfish nodes in the network and isolates them from the network activities so that energy of the nodes and the precious bandwidth of the wireless links are optimally utilized. The second protocol is based on an algorithm for detection of selfish nodes in a WMN that uses statistical theory of inference and clustering techniques to make a robust and reliable classification of the nodes based on their packet forwarding activities. It also introduces some additional fields in the packet header for AODV protocol so that detection accuracy is increased. In the following sub-sections the two protocols are discussed in detail.

## 5.1 A secure and efficient routing protocol for WMNs

A secure and efficient routing protocol for WMNs has been proposed in (Sen, 2010a) that can handle stringent quality of service (QoS) requirements of real-time applications. There are several key contributions of the work: (i) It provides an accurate estimation of the end-to-end delay in a routing path; the estimated value is then used to check whether the routing can guarantee the application QoS. (ii) It computes a link quality estimator and utilizes it in route selection. (iii) It provides a framework for reliable estimation of available bandwidth in a routing path so that flow admission with guaranteed QoS can be made. (iv) It helps in identifying and isolating selfish nodes.

The protocol is a reactive routing protocol, in which during the routing discovery phase, each intermediate node uses an admission control scheme to check whether the flow can be admitted or not. If a flow is admitted, an entry is created for the flow in a table (called the flow table) maintained locally by the node. The important components of the protocol are described below:

i.   **Estimating reliability of routing paths:** every node estimates the reliability of each of its wireless links to its one-hop neighbor nodes. For computing the reliability of a link, the number of control packets that a node receives in a given time window is used as a base parameter. An *exponentially weighted moving average* (EWMA) method is used to update the link reliability estimate. If the percentage of control packets received by a node over a link in the last interval of measurement of link reliability is $N_t$, and if $N_{t-1}$ is the historical value of the link reliability before the last measurement interval, $\alpha = 0.5$ is the weighting parameter, the updated link reliability ($R$) is computed using (8):

$$R = \alpha * N_t + (1 - \alpha) * N_{t-1} \tag{8}$$

Every node maintains estimates of the reliability of each of its links with its neighbors in a *link reliability table*. The reliability for an end-to-end routing path is computed by taking the average of the reliability values of all the links on the path. Computation of the link reliability values is based on the RREQ packets on the reverse path and the RREP packets on the forward path. The use of routing path with the highest reliability reduces the overhead of route repair and makes the routing process more efficient.

ii.  **Use of network topological information in route discovery:** the protocol makes use of the knowledge of network topology by utilizing selective flooding of control messages in a portion of the network. In this way, broadcasting of control messages is avoided and thus the chances of network congestion and disruption of the flows in the network are reduced. If both the source and the destination are under the control of the same mesh router (Fig. 15), the flooding of the control messages are confined within the portion of the network served by the mesh router only. However, if the source and the destination are under different mesh routers, the control traffic is limited to the two mesh groups. To reduce the control overhead further and enhance the routing efficiency, the nodes accept broadcast control messages from only those neighbors which have link reliability greater than 0.5 (i.e., on the average 50% of the control packets sent from those nodes have been received by the node). This ensures that paths with less reliability are not discovered, and hence not considered for routing.

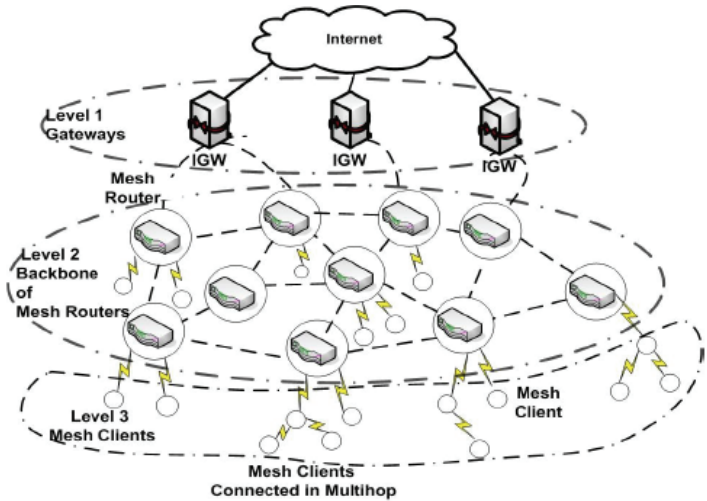

Fig. 15. The hierarchical architecture of a WMN

iii. **Estimating end-to-end delay in a routing path:** for addressing the issue of differential delays experienced by the control and data packets, the protocol makes use of some *probe packets* during the route discovery phase. When a source node receives RREP packets from the destination in response to its RREQ, it stores in a table, the records for all the RREP packets together with the path through which the packets have arrived at it. However, instead of randomly selecting a path to send probe packets to the destination, the packets are sent along the path from which RREP messages have arrived at the source first. This ensures that the probe packets are sent along the path which is likely to induce less end-to-end delay, resulting in a better performance of the protocol. The probe packets are identical to the data packets so far as their size, priority, and flow rates are concerned. The objective of sending probe packets is to simulate the data flow and observe the delay characteristics in the routing path. The number of probe packets is kept limited to $2H$ for a path consisting of $H$ hops to make a trade-off between the control overhead and measurement accuracy (Kone et al., 2007). The destination node sets a timer after it receives the first probe packet from the source node. The timer duration is based on the estimated time for receiving all the probe packets and is computed statistically. The destination computes the average delay experienced by all the probe packets it has received, and send the computed value to the source node piggybacking it on an RREP message. If the computed value is within the limit of tolerance of the application QoS, the source selects the route and sends packets through it. If the delay exceeds the acceptable limit, the source selects the next best path (based on the arrival of RREP packets) from its table and tries once again. Since the routing path is set up based on probe packets rather than the naïve RREP packets, the protocol has higher route establishment time. However, since the selected paths have high end-to-end reliability, the delay and the control overhead are reduced because of minimal subsequent route breaks.

iv. **Estimation of available network bandwidth:** the protocol estimates the available bandwidth in a wireless link using its end-to-end delay and the loss of packets due to congestion. The packet loss due to congestion in the link is estimated as follows. In a wireless link packet loss may happen due to tow reasons: (a) loss due to faulty wireless links and (b) loss due to network congestion. The *radio link control* (RLC) layer segments an IP packet into several RLC frames before transmission and reassembles them into an IP packet at the receiver side. An IP packet loss occurs when any RLC frame belonging to an IP packet fails to be delivered. When this happens, the receiver knows that the RLC frames re-assembly has failed and the IP packet has been lost due to wireless error. Meanwhile, the sender detects *retransmission time out* (RTO) of the frame and discards all the RLC frames belonging to the IP packet. This enables the sender to compute packet drop rate in the wireless links. Moreover, using the sequence numbers of the IP packets received at the receiver, it is possible to differentiate the packet loss due to link error and packet loss due to congestion (Yang et al., 2004). For example, while receiving two incoming packets with sequence number $i$ and $i$ +2, if the receiver finds an IP packet assembly failure in RLC layer, the packet with sequence number $i$+1 is lost due to wireless channel. Once the packet loss ratio due to congestion ($P_{congestion}$) is estimated, the available bandwidth in the wireless link, *estrat*, is computed as follows (Yang et al., 2004):

$$estrat = \frac{PacketSize}{X + Y} \qquad (9)$$

In (9), $X$ and $Y$ are given by:

$$X = RTT\sqrt{\frac{2Pcongestion}{3}} \qquad (10)$$

$$Y = RTO * \min(1, 3 * \sqrt{\frac{3Pcongestion}{8}}Pcongestion(1 + 32P_{congestion}^2) \qquad (11)$$

In (10), RTT is the average round trip time for a control packet. RTO is the retransmission time out for a packet, and is computed using (12):

$$RTO = \overline{RTT} + k * \overline{RTT}_{Var} \qquad (12)$$

In (12), $\overline{RTT}$ and $\overline{RTT}_{Var}$ are the mean and variance respectively of RTTs and $k$ is set to 4. This bandwidth estimator is employed to dynamically compute the available bandwidth in the wireless links on a routing path so that the guaranteed minimum bandwidth for the flow is always maintained throughout the application life-time.

**v.   Identifying selfish nodes:** the protocol also enforces cooperation among the nodes by identifying the selfish nodes in the network and isolating them. Selfishness is an inherent problem associated with any capacity-constrained multi-hop wireless networks like WMNs. A mesh router can behave selfishly owing to various reasons such as: (a) to obtain more wireless or Internet throughput, or (b) to avoid path congestion. A selfish mesh router increases the packet delivery latency, and also increases the packet loss rate. A selfish node while utilizing the network resources for routing its own packet, avoids forwarding packets for others to conserve its energy.

Identification of selfish nodes is therefore, a vital issue. Several schemes have been proposed in the literature to mitigate the selfish behavior of nodes in wireless networks, such as credit-based schemes, reputation-based schemes, and game theory-based scheme (Santhanam et al., 2008). However, to keep the overhead of computation and communication at the minimum, the protocol employs a simple mechanism to discourage selfish behavior and encourage cooperation among nodes. To punish the selfish nodes, each node forwards packets to its neighbor node for routing only if the link reliability of the latter is greater than a threshold value (say, 0.5). Since the link reliability of a selfish node is 0, the packets arriving from this node will not be forwarded. Therefore, to keep link reliability higher than the threshold, each node has to participate and cooperate in routing. The link reliability serves dual purpose of enhancing reliability and enforcing node cooperation in the network.

vi. **QoS violation and recovery:** the protocol detects failure to guarantee QoS along a path with the help of reservation timeouts in flow tables records maintained in the nodes, by detection of non-availability of minimum bandwidth as estimated along its outbound wireless link. Failure to guarantee QoS may occur in three different scenarios. In the first case, a node receives a data packet for which it does not find a corresponding record in its flow table. This implies that a reservation time-out has happened for that flow. The node, therefore, sends a *route error* (RERR), to the source which re-initiates route discovery. In the second scenario, a destination node detects from its flow table records that the data packets received have exceeded the maximum allowable delay ($T_{max}$). To restore the path, the destination broadcasts a new RREP back to the source, and the source starts re-routing the packets via the same path on which RREP has traversed. In the third case, an intermediate node on the routing path may find that the estimated bandwidth (using (9)) in its forwarding link is less than the guaranteed minimum ($B_{min}$) value. In this case, the intermediate node sends an RERR to the source which re-initiates the route discovery process. The real-time estimation of the bandwidth in the next-hop wireless link at each node on the routing path makes the protocol more robust and reliable compared to most of the existing routing protocols for WMNs. For example, the similar protocol presented in (Kone et al., 2007) does not employ any bandwidth estimation mechanism at intermediate nodes, and therefore, cannot ensure delivery of all packets for every admitted flow in the network.

## 5.2 A Trust-based protocol for selfish nodes detection in WMNs

To address the issue of selfish nodes in a WMN, a scheme has been proposed that uses local observations in the nodes for detecting node misbehavior (Sen, 2010b). The scheme is applicable for on-demand routing protocol like *ad hoc on-demand distance vector* (AODV) protocol, and uses statistical theory of inference and clustering techniques to make a robust and reliable classification (cooperative or selfish) of the nodes based on their neighbors. In addition, the scheme introduces additional fields in the packet header of AODV packets so that detection accuracy is increased. Since the security protocol works on AODV protocol, a brief description of AODV protocol is given before the protocol is described for the benefit of the readers.

**AODV protocol and modeling of the state machine:** AODV routing protocol uses an on-demand approach for finding routes to a destination node. It employs destination sequence numbers to identify the most recent path. The source node and the intermediate nodes store the next-hop information corresponding to each flow of data packet transmission. The source node floods the *route request* (RREQ) packet in the network when a route is not available for the desired destination. It may obtain multiple routes to different destinations from a single RREQ. The RREQ carries the source identifier (*src_id*), the destination identifier (*dest_id*), the source sequence number (*src_seq_num*), the destination sequence number (*dest_seq_num*), the broadcast identifier (*bcast_id*), and the *time to live* (TTL). When an intermediate node receives an RREQ, it either forwards the request further or prepares a *route reply* (RREP) if it has a valid route to the destination. Every intermediate node, while forwarding an RREQ, enters the previous node address and its *bcast_id*. A timer is used to delete this entry in case an RREP is not received before the timer expires. This helps in storing an active path at the intermediate node as AODV does not employ source routing of data packets. When a node receives an RREP packet, information of the previous node from which the packet was received is also stored, so that data packets may be routed to that node as the next hop towards the destination. It is clear that AODV depends heavily on cooperation among the nodes for its successful operation. A selfish node can easily manipulate the protocol to minimize its chances of being included on routes for it is neither the source nor the destination. It may drop or tamper with the RREQ messages to ensure that no routes will ever be selected through it. Alternatively, it may drop, delay, or modify the RREP messages so as to prevent the replies from reaching the source node. The security protocol proposed in this work attempts to detect selfish nodes in a WMN so that these nodes may be isolated from the network. In the following, a *finite state machine* (FSM) model of the AODV protocol is presented which is utilized later for describing the security protocol.

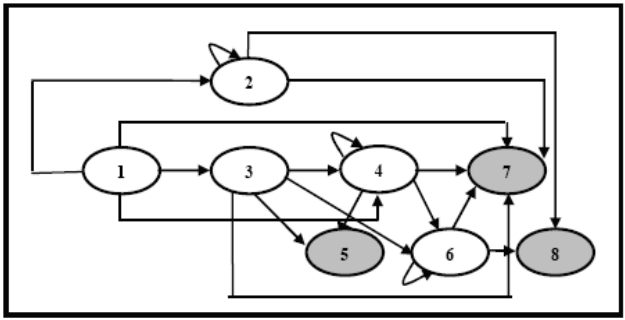

Fig. 16. The finite state machine of a monitored node

**Finite state machine model:** in the security mechanism, with AODV as the underlying routing protocol, the set of all messages corresponding to a RREQ flooding and the unicast RREP is referred to as a *message unit*. It is clear that no node in the network can observe all the transmission in a message unit. The subset of a message unit that a node can observe is referred to as the *local message unit* (LMU). The LMU for a particular node consists of the messages transmitted by that node, the messages transmitted by all its neighbors, and messages overheard by the node. The detection of selfish nodes is made on the basis of data collected by each node from its observed LMUs. Corresponding to each message transmission in an LMU, a node maintains a record of its sender, and the receiver in its neighborhood. It also keeps record of the neighbor nodes that receive the RREQ broadcast messages sent by the node itself. The messages are assumed to follow the sequence of the AODV protocol. The finite state machine shown in Fig. 16 depicts various states through which a neighbor node undergoes for each LMU (Wang et al., 2008). The corresponding states for the numbers mentioned in Fig.16 can be found in Table 2.

| State | Interpretation |
|-------|----------------|
| 1: init | Initial phase; no RREQ is observed |
| 2: unexp RREP | Receipt of a  RREP without RREQ observed |
| 3: rcvd RREQ | Receipt of a RREQ observed |
| 4: fwd RREQ | Broadcast of a RREQ observed |
| 5: timeout RREQ | Timeout after receipt of RREQ |
| 6: rcvd RREP | Receipt of a RREP observed |
| 7: LMU complete | Forwarding of a valid a RREP observed |
| 8: timeout RREP | Timeout after receipt of a RREP |

Table 2. The states of the finite state machine for a local message unit (LMU)

To distinguish the finals states, these states are shaded.  Every message transmission by a node causes a state transition in each of its neighbor's finite state machine. The finite state machine in one neighbor node gives only a local view of the activities of the node being monitored. It does not, in any way, represent the actual behavior of the monitored node. The collaborative participation of each neighbor node makes it possible to get an accurate global picture regarding the monitored node's behavior. A node whose activity is being monitored by its neighbors is referred to as a *monitored node*, and its neighbors are referred to as a *monitor node*. Each node plays the dual role of a monitor node and a monitored node for each of its neighbors. Each monitor node in the network observes a series of interleaved LMUs for a routing session. Each LMU can be identified by the source-destination pair contained in an RREQ message. Let us denote the $k$th LMU observed by a monitor node as $(s_k, d_k)$. The pair $(s_k, d_k)$ does not uniquely identify an LMU, because source can issue multiple RREQs for the same destination. However, since the subsequent RREQs have some delays associated with them, we can safely assume that there is only one active LMU $(s_k, d_k)$ in the network at any point of time. At the beginning, a monitored node starts with the state 1 in its finite state machine. As the monitor node(s) observes the behavior of the monitored node by examining the LMUs, it records a sequence of transitions form its initial state 1 to one of its possible final states -- 5, 7 and 8. When a monitor node broadcasts an RREQ, it assumes that the monitored node has received it. The monitor node, therefore, records a state transition 1 → 3 for the monitored node's finite state machine. If a monitor node observes a monitored node to broadcast an RREQ, then a state transition of 3 → 4 is recorded if the RREQ message was previously sent by the monitor node to the monitored node; otherwise a transition of 1 → 4 will be recorded meaning thereby that the RREQ was received by the monitored node from some other neighbor. The transition to a timeout state occurs when a monitor node finds no activity by the monitored node for the concerned LMU before the expiry of a timer. When a monitor node observes a monitored node to forward an RREP, it records a transition to the final state – *LMU complete* (State No 7). At this state, the monitored node becomes a candidate for inclusion on a routing path.

Fig. 17 depicts an example of LMU observed by the node *N* during the discovery of a route from the source node *S* to the destination node *D* indicated by bold lines. Table 3 shows the events observed by node *N* and the corresponding state transitions for each of its three neighbor nodes *X*, *Y* and *Z*. When the final state is reached, the finite state machine terminates and the corresponding sequences of state transitions are stored by each node for each of its neighbors. When sufficient number of events is collected by a node, a statistical analysis is performed to detect the presence of any selfish nodes in the network.

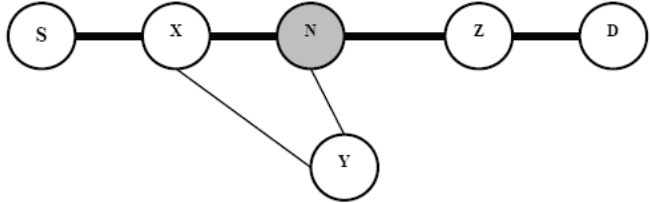

Fig. 17. An example of local message unit (LMU) observed by node *N*

| Neighbor | Events | State changes |
|---|---|---|
| *X* | *X* broadcasts RREQ | 1 → 4 |
| | *N* broadcasts RREQ | 4 → 4 |
| | *N* sends RREP to *X* | 4 → 6 |
| | *X* sends RREP to *S* (overheard) | 6 → 7 |
| *Y* | *Y* broadcasts RREQ | 1 → 4 |
| | *N* broadcasts RREQ | 4 → 4 |
| | Timeout | 4 → 5 |
| *Z* | *N* broadcasts RREQ | 1 → 3 |
| | *Z* broadcasts RREQ | 3 → 4 |
| | *Z* sends RREP to *N* | 4 → 7 |

Table 3. The state transitions of the neighbor nodes of node *N*

**The security algorithm:** As mentioned in the previous section, a monitoring node keeps a record of state transitions in the finite state machine of a monitored node in each LMU. These sequences can be represented as a transition matrix $T = [T_{ij}]$, where $T_{ij}$ is the number of times the transition $i \rightarrow j$ is found. The monitor node invokes a detection algorithm every *W* seconds using data from the most recent $D = d * W$ seconds of observations, where *d* is a small integer. The parameter *D*, called the *detection window*, should be such that it is possible to punish the selfish nodes promptly while maintaining a high level of accuracy.

In the proposed algorithm, a node is assumed to monitor the activities of its *R* neighbors which are identified by their respective indices 1, 2,….*R*. Let $T^{(r)} = \left[ f_{ij}^{(r)} \right]$ denote the observed transition matrix for the *r*th neighbor, where $[f_{ij}^{(r)}]$ is the number of transitions from state *i* to state *j* observed in the previous detection window. If *m* is the number of states in the finite state machine in each node, the size of $T^{(r)}$ is *m* x *m*. Let $T^{(r)} = [f_{i1}^{(r)}, ... f_{im}^{(r)}]$ denote the *i*th row of the transition matrix $T^{(r)}$, which shows the transitions out of state *i* at the neighbor node *r*. If two neighbor nodes *r* and *s* have identical distributions corresponding to transitions from state i, then one can write $T_i^{(r)} \equiv T_i^{(s)}$.

To test the hypothesis $T_i^{(r)} \equiv T_i^{(s)}$ the Pearson's χ2 test is used as follows.

$$\chi^2(i) = \frac{\sum\limits_{l\varepsilon(r,s)} \sum\limits_{j=1}^{m} \left[ f_{ij}^{(l)} - \overline{f}_{ij}^{(l)} \right]^2}{\overline{f}_{ij}^{(l)}} \tag{13}$$

$$\overline{f}_{ij}^{(l)} = F_{ij}^{(l)} \frac{f_{ij}^{(r)} + f_{ij}^{(s)}}{F_i^{(r)} + F_i^{(s)}} \tag{14}$$

where $F_i^{(r)}$ and $F_i^{(s)}$ denote total number of transitions for state $i$ in $T^{(r)}$ and $T^{(s)}$ respectively.

If the value of $\chi^2$ exceeds the value of $\chi^2_{m-1,\alpha}$, then the hypothesis $T_i^{(r)} \equiv T_i^{(s)}$ is rejected at confidence interval $\alpha$. If we write $K_i^{rs}$ for the event that $\chi^2_{(i)} > \chi^2_{m-1,\alpha}$, then the conditional probability $P\left(T_i^{(r)} \equiv T_i^{(s)} \mid B_i^{rs}\right)$ can be taken as a reasonable estimator of the similarity between $r$ and $s$ with respect to the state $i$. In absence of any prior information, it is reasonable to assume that $r$ and $s$ have no similarity in state $i$ and the probability that the Pearson test rejects its hypothesis to be 0.5 (Wang et al., 2008). In order to evaluate the similarity between $r$ and $s$ for all the m states, (1) is applied to all rows of $T(r)$ and $T(s)$. This yields a vector, $\{i = 1,.....,m\}$. From the standard Markovian principle one can write:

$$L_{rs} = P(T^{(r)} \equiv T^{(s)} \mid B^{(rs)})$$
$$= \alpha^{S^{(rs)}} (1-\alpha)^{m-S^{(rs)}} \approx \alpha^{S^{(rs)}} \tag{14}$$

where $$S^{(rs)} = \sum_{i=1}^{m} B_i^{(rs)} \tag{15}$$

The lower-order terms in the right hand side of (15) are ignored since $a \ll 1$. For small value of $a$, $L_{rs}$ monotonically decreases in $S^{(rs)}$, which, as evident from (15), is the number of rejections of Pearson's hypothesis. Therefore, $1 -- L_{rs}$ may be taken as the measure of the dissimilarity between the neighbor nodes $r$ and $s$. In presence of noise in the data, however, it is found that for two nodes $r$ and $s$ which have $L_{rs} \approx 1$, a third node $t$ may cause inconsistency such that $L_{rt} \neq L_{st}$. To avoid this inconsistency in clustering in the proposed algorithm, clustering are not computed on the basis of pair-wise dissimilarity. To compute dissimilarity between $r$ and $s$, the $L$ values for all neighbors are computed with respect to $r$ and $s$ separately, and the following equation is applied:

$$d_{rs} = 1 - \frac{n_{rs}^2}{n_{r/s} * n_{s/r}} \tag{16}$$

where,

$$n_{rs} = \sum_{t \neq r,s} \min(L_{rt}, L_{st}),$$

$$n_{r/s} = \sum_{t \neq r,s}^{K} L_{rt}$$

$$n_{s/r} = \sum_{t \neq r, s}^{K} L_{st}$$

It may be observed that the computation of $d_{rs}$ does not involve $L_{rs}$ -- the pair-wise similarity index between nodes $r$ and $s$. In fact, it measures the degree of inconsistency in similarity between $r$ and $s$ with all their neighbors. Since, in the computation, contribution of each neighbor plays its role, $d_{rs}$ presents a robust indicator for dissimilarity between nodes and plays a crucial part in computing the clusters (Wang et al., 2008). For clustering, an *agglomerative hierarchical clustering* technique is used. This is a single-linkage approach in which each cluster is represented by all of the objects in the cluster, and the similarity between two clusters is measured by the similarity of the closest pair of data points belonging to different clusters. The cluster merging process repeats until all the objects are eventually merged to form one cluster (Eddy et al., 1996). After the nodes are clustered into similar sets, the sets are further classified into three groups: (i) a set ($G$) of cooperative nodes, (ii) a set ($B$) of selfish nodes, and (iii) a set of nodes whose behavior could not be ascertained. The cooperation score ($C_r$) of a node is computed as (Wang et al., 2008):

$$C_r = \frac{\sum_{i,j \varepsilon G}^{m} n_{ij}^{(r)}}{|G|} - \frac{\sum_{i,j \varepsilon B}^{m} n_{ij}^{(r)}}{|B|} \tag{17}$$

The set $B$ is most likely to contain the selfish nodes. To reduce false positives (i.e. wrongly identifying a cooperative node as selfish), an ANOVA test is applied. The ANOVA approach computes a probability $P_k$ of the random variation among the mean cooperation scores of $k$ clusters. A lower value of $P_k$ implies that the clusters actually represent distinct differences in their behavior. At each iteration, $k$ clusters are formed and $P_k$ is compared with a pre-defined level of significance β. If $P_k$ < β, clusters are believed to be reliably reflecting the behavior of the nodes and their classifications are accepted. The cluster with lowest mean cooperation score is assumed to contain the selfish nodes. If $P_k$ > $P_{k-1}$, the neighbor behavior has not been properly reflected in the cluster formation, which has led to the increase in the value of $P_k$. In this case, all the nodes are classified as cooperative, and the next iteration of the algorithm is executed. The confidence parameter β can be tuned so as to adjust the alacrity of detection of selfish nodes and rate of false positives (Wang et al., 2008). In spite of all the above statistical approaches, there is still a possibility of misclassification. The proposed algorithm further reduces the probability of misclassification by a new cross-checking mechanism. For this purpose, a minor modification is suggested in the packet header for AODV routing. Two additional fields are inserted in the header of an RREQ packet. These fields are: *next_to_source* and *duplicate_flag* to indicate respectively the address of the node that is next hop to the source, and whether the packet is a duplicate packet which has already been broadcasted by some other nodes in the network. In the header of an RREQ packet, in addition to the above two fields, another field called *next_to_destination* is added to indicate the address of the node to which the packet must be forwarded in the reverse path. It has been shown in (Kim et al., 2008), with the above extra fields, it is possible to detect every instance of selfish behavior in a wireless network with 100% detection accuracy, if the following conditions are satisfied: (i) no packet loss lost due to interference, (ii) links are bi-directional, (iii) the nodes are stationary, and (iv) the queuing delays are bounded. Since all these conditions cannot be guaranteed in a real-world deployment, there will be always some detection inaccuracy.

Table 4 presents a list of vulnerabilities in different layers of the protocol stack of WMNs and the security protocols for defending those attacks. Table 5 compares the secure routing protocols discussed in this chapter with respect to various mechanisms these protocols use.

## 6. Conclusion

WMNs have become the focus of research in recent years, owing to their great promise in realizing numerous next-generation wireless services. Driven by the demand for rich and high-speed content access, recent research on WMNs has focussed on developing high performance communication protocols, while the security of the proposed protocols have received relatively little attention. However, given the wireless and multi-hop nature of the communication, WMNs are subject to a wide range of security threats. In this chapter, a large number of security issues at various layers of WMNs have been presented with a particular focus on the network layer. In addition, some of the major routing security mechanisms for WMNs currently existing in the literature have been presented and compared with respect to their strengths and weaknesses. A few novel secure routing mechanisms that take into account application QoS while detecting malicious and selfish nodes are also discussed. Although, researchers have done substantial contributions in the area of routing security in WMNs, there are still many challenges that remain to be addressed. First, efficient (i.e., lightweight) and robust authentication protocols for the *mesh routers* (MRs) need to be designed which involves scalable key management techniques. Second, for reliability in routing, energy-aware and secure multi-path routing protocols are in demand. Third issue is on strategic deployment of hop integrity protocols in WMNs. Hop integrity protocols are open to incremental deployment, and the security they provide increases with the number of pairs of hop integrity-equipped mesh routers, because an adversary will have less venues to launch his/her attacks. However, due to hardware/software compatibility and efficiency consideration, it may be worthwhile to consider a strategic deployment scheme. For example, few hotspots in the network may be required to install static hop integrity, in which hop integrity is always turned on; other spots in the network can install dynamic hop integrity, in which hop integrity is randomly turned on and off. Fourth, efficient security mechanism should be designed for defending against *tunnelling* attack, in which two malicious nodes advertise in such a way as if they have a very reliable link between them. This is achieved by tunnelling AODV messages between them. No security scheme exists so far that can detect this attack promptly and efficiently. Fifth, appropriate security protocols should be designed for hybrid networks. In many deployment situations, WMNs are designed to be integrated with other types of networks, such as wired networks and cellular networks. Addressing attacks in hybrid environment also presents an interesting future direction. Such networks are vulnerable to a wider range of attacks than its individual network components. For example, a mesh network for wireless Internet access can be targeted with DDoS attacks launched from the Internet. The scarcity of bandwidth resource on WMNs further exacerbates the severity of such attacks. On the other hand, hybrid networks possess additional resources and opportunities for defending against attacks. For example, WMNs connected to the wired networks, it is possible to leverage the high bandwidth, low latency wired links, and deploy powerful computers on the wired networks to defend against attacks. Sixth, a balanced network coding system needs to be designed for high performance secure routing (Ahlswede et al., 2000). Existing network coding systems are vulnerable to a wide range of attacks besides the most well-known *packet pollution attacks* (Yu et al., 2008). Many of the weaknesses of existing system designs lie in their single focus in performance optimizations. A more balanced approach, which can provide improved security guarantees, is crucial for the actual adoption of network coding in real-world applications. A future direction of research is to uncover the security implications of different design and optimization techniques, and explore balanced system designs with network coding that achieve appropriate tradeoffs between security and performance suitable for different application requirements. Finally, multi-layer (i.e. cross-layer) security protocols should be developed that address network vulnerabilities in multiple layers of the protocol stack to provide robust and highest level of protection to mission-critical network deployments.

| Attack | Targeted layer in the protocol stack | Protocols |
|---|---|---|
| Jamming | Physical and MAC layers | Frequenscy hopping spread spectrum (FHSS), Direct sequence spread spectrum (DSSS) |
| Wormhole | Network layer | Packet Leashes (Hu, 2003b) |
| Blackhole | Network layer | SAR (Yi, 2001) |
| Grayhole | Network layer | GRAYSEC (Sen, 2007), SAR (Yi, 2001) |
| Sybil | Network layer | SYIBSEC (Newsome, 2004) |
| Selective packet dropping | Network layer | SMT (Papadimitratos, 2003a), ARIADNE (Hu, 2002a), Sen (2010a), Sen (2010b) |
| Rushing | Network layer | ARAN (Sanzgiri, 2002), SAR (Yi, 2001), SEAD (Hu, 2002b), ARIADNE (Hu, 2002a), SAODV (Li, 2001), SRP (Papadimitratos, 2002), SEAODV (Li, 2011) |
| Byzantine | Network layer | ODSBR (Awerbuch, 2002) |
| Resource depletion | Network layer | SEAD (Hu, 2002b) |
| Information disclosure | Network layer | SMT (Papadimitratos, 2003a) |
| Location disclosure | Network layer | SRP (Papadimitratos, 2002) |
| Routing table modification | Network layer | ARAN (Sanzgiri, 2002), SAR (Yi, 2001), SRP (Papadimitratos, 2002), SEAD (Hu, 2002b), ARIADNE (Hu, 2002a), SAODV (Li, 2001), SEAODV (Li, 2011) |
| Repudiation | Application layer | ARAN (Sanzgiri, 2002) |
| Denial of service | Multi-layer | SRP (Papadimitratos, 2002), SEAD (Hu, 2002b), ARIADNE (Hu, 2002a) |
| Impersonation | Multi-layer | ARAN (Sanzgiri, 2002), SEAD (Hu, 2002b), SEAODV (Li, 2011) |

Table 4. Different attacks on WMN protocol stack and protocols for defending the attacks

| Protocol | Secret Keys | MAC | Digital Signature | Hash Chain | Cryptographic mechanism | Assumptions | Verification mechanism |
|---|---|---|---|---|---|---|---|
| ARAN (Sanzgiri, 2002) | Public and private key pair for each node | ------ | ------ | ------ | Public key cryptography | Trusted certificate server | Public key cryptography verification mechanism |
| SEAD (Hu, 2003) | Initial secret key Ks for hash function | ------ | ------ | Authenticates the sequence number and routing table metric by one-way hash chain | ------ | Secure way of delivering initial secret | Hash chain verification |
| SAR (Yi, 2001) | Public and private key pairs for each node | ------ | ------ | ------ | Public key cryptography | Trust metric computation model is present in the network | Trust verification at each node on routing path |
| SAODV (Zapata, 2002) | Public and private key pair for each node | ------ | Sender uses digital signature to sign the messages | One-way hash chain to authenticate hop-count | ------ | Network should have a key distribution system | Digital signature verification system |
| SRP (Papadimratos, 2002) | SA between source and destination | MAC computation with Key | ------ | ------ | ------ | Secure ware of delivering the SA | MAC verification mechanism |
| ARIADNE (Hu, 2002) | Secret MAC keys between sender and receiver | MAC also | ------ | TESLA keys authenticate message. It uses hash-chain to generate these keys | ------ | Nodes have loosely synchronized clocks | MAC verification mechanism |
| SEAODV (Li, 2011) | Pairwise transient keys and group transient keys | Broadcast message: MACGtk; Unicast message: MACPtk | ------ | ------ | Symmetric key cryptography | Bloom's key pre-distribution system is present | MAC verification mechanism |
| SECROUTE (Sen, 2010a) | Public and private key pair for each node | ------ | ------ | ------ | Public key cryptography | Nodes have watchdog mechanisms | Link reliability verification by packet forwarding statistics analysis |
| TRUSTROUTE (Sen, 2010b) | Public and private key pair for each node | ------ | ------ | ------ | ------ | Nodes have watchdog mechanisms | Agglomerative clustering mechanism to identify selfish nodes |

Table 5. Comparative analysis of various secure routing protocols for WMNs